\documentclass[journal]{IEEEtran}
\usepackage{amsmath,amsfonts}
\usepackage{algorithmic}
\usepackage{algorithm}
\usepackage{array}
\usepackage[caption=false,font=normalsize,labelfont=sf,textfont=sf]{subfig}
\usepackage{textcomp}
\usepackage{stfloats}
\usepackage{url}
\usepackage{verbatim}
\usepackage{graphicx}
\usepackage{cite}
\usepackage{hyperref}
\usepackage{booktabs} 
\usepackage{tabularray}
\usepackage{multirow}
\usepackage[normalem]{ulem}
\usepackage{xcolor}
\usepackage{makecell}
\usepackage{tikz}
\usepackage{cuted}
\usepackage{capt-of}
\usepackage{mathptmx}
\usepackage{amssymb}
\usepackage{pdfrender}
\usetikzlibrary{shadows,fit, backgrounds}
\usepackage{float}
\usepackage{mwe}
\graphicspath{{figures/}{pictures/}{images/}{./}} 
\usepackage{xstring}
\usepackage{tabu}
\usepackage{lipsum} 
\usepackage{tabularx}
\usepackage{CJKutf8}
\usepackage{xspace}

\definecolor{tagred}{HTML}{7E0900}
\definecolor{tagcream}{HTML}{F2ECDB}

\makeatletter
\DeclareRobustCommand{\paneltag}[1]{%
  \StrLen{#1}[\paneltag@len]%
  \ifnum\paneltag@len=1
    \def\paneltag@font{\rmfamily\bfseries\fontsize{7pt}{7pt}\selectfont}%
  \else
    \def\paneltag@font{\rmfamily\bfseries\fontsize{5pt}{5pt}\selectfont}%
  \fi
  \raisebox{0.12ex}{%
    \tikz[baseline=(char.base)]{
      \node[
        rounded corners=0.75pt,
        fill=tagred,
        minimum size=3.5mm,
        inner sep=0pt,
      ] (outer) {};
      \node[
        rounded corners=0.6pt,
        fill=tagcream,
        minimum size=3.0mm,
        inner sep=0pt,
      ] (frame) at (outer.center) {};
      \node[
        rounded corners=0.45pt,
        fill=tagred,
        text=white,
        font=\paneltag@font,
        minimum size=2.35mm,
        inner sep=0pt,
      ] (char) at (outer.center) {#1};
    }%
  }%
}
\makeatother

\definecolor{panelTagRed}{HTML}{9A3A33}

\DeclareRobustCommand{\paneltagRound}[1]{%
  \raisebox{0.32ex}{%
    \scalebox{0.60}{%
      \tikz[baseline=(char.base)]{
        \node[
          circle,
          fill=panelTagRed,
          text=white,
          font=\rmfamily\bfseries\fontsize{11pt}{12pt}\selectfont,
          minimum size=5.2mm,
          inner sep=0pt,
          drop shadow={shadow xshift=0.25pt, shadow yshift=-0.25pt, opacity=0.25}
        ] (char) {#1};
      }%
    }%
  }%
}

\DeclareRobustCommand{\syscomp}[1]{%
\tikz[baseline=(X.base)]\node[
    fill=black!8,
    rounded corners=2.5pt,
    inner xsep=4.5pt,
    inner ysep=2pt,
    font=\normalsize \itshape
](X){#1};%
}

\DeclareRobustCommand{\syscompInCaption}[1]{%
  \tikz[baseline=(X.base)]%
  \node[
    fill=black!8,
    rounded corners=2pt,
    inner xsep=2.5pt,
    inner ysep=0.6pt,
    font=\footnotesize\itshape
  ](X){#1};%
}

\newcommand{\entityText}[2]{\textcolor{#1}{\textbf{\textit{#2}}}}
\newcommand{\toolname}{\textbf{\textit{CompoVista}}\xspace}

\newcommand{\TCP}{\textbf{TCP}\xspace}
\newcommand{\compograph}{\textbf{\textit{CompoGraph}}}

\definecolor{scopeHuman}{HTML}{703434}
\definecolor{scopeVegetation}{HTML}{1F502A}
\definecolor{scopeLandform}{HTML}{005252}
\definecolor{scopeWater}{HTML}{004E71}
\definecolor{scopeAnimal}{HTML}{5B4218}
\definecolor{scopeArtefact}{HTML}{523E6D}
\definecolor{scopeOtherEntity}{HTML}{8B95A7}

\definecolor{scopeEntity}{HTML}{000000}
\definecolor{scopeContext}{HTML}{000000}
\definecolor{scopeRelation}{HTML}{000000}
\definecolor{scopeVoid}{HTML}{000000}

\NewDocumentCommand{\entityIconText}{O{1.0em} O{-0.30ex} O{0.12em} m m m}{%
  \mbox{%
    \raisebox{#2}{\includegraphics[height=#1]{#4}}%
    \hspace{#3}\entityText{#5}{#6}%
  }%
}

\NewDocumentCommand{\entityHumanTagText}{O{1.1em} O{-0.30ex} O{0.15em} m}{%
  \entityIconText[#1][#2][#3]{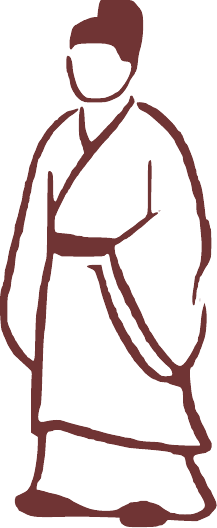}{scopeHuman}{#4}%
}

\NewDocumentCommand{\entityVegetationTagText}{O{1.1em} O{-0.30ex} O{0.12em} m}{%
  \entityIconText[#1][#2][#3]{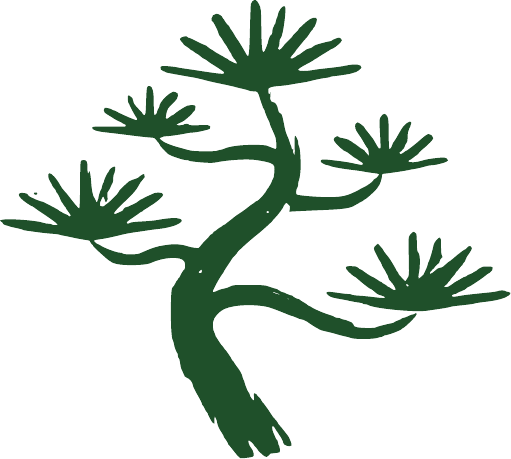}{scopeVegetation}{#4}%
}

\NewDocumentCommand{\entityLandformTagText}{O{1.0em} O{-0.30ex} O{0.12em} m}{%
  \entityIconText[#1][#2][#3]{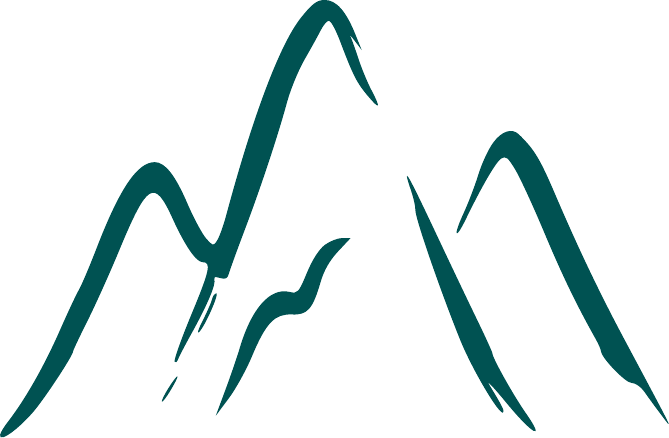}{scopeLandform}{#4}%
}

\NewDocumentCommand{\entityWaterTagText}{O{0.9em} O{-0.30ex} O{0.12em} m}{%
  \entityIconText[#1][#2][#3]{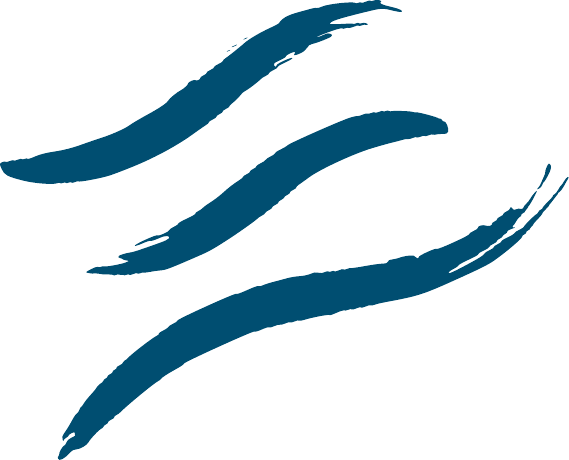}{scopeWater}{#4}%
}

\NewDocumentCommand{\entityAnimalTagText}{O{1.0em} O{-0.30ex} O{0.12em} m}{%
  \entityIconText[#1][#2][#3]{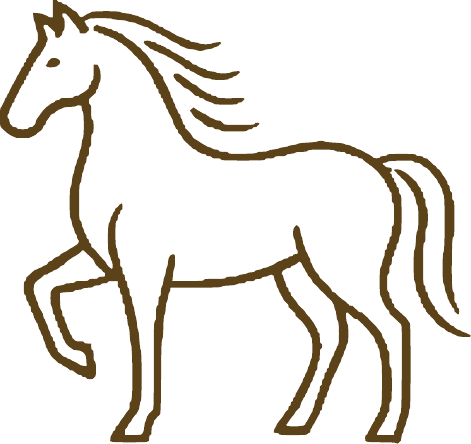}{scopeAnimal}{#4}%
}

\NewDocumentCommand{\entityArtefactTagText}{O{1.0em} O{-0.30ex} O{0.12em} m}{%
  \entityIconText[#1][#2][#3]{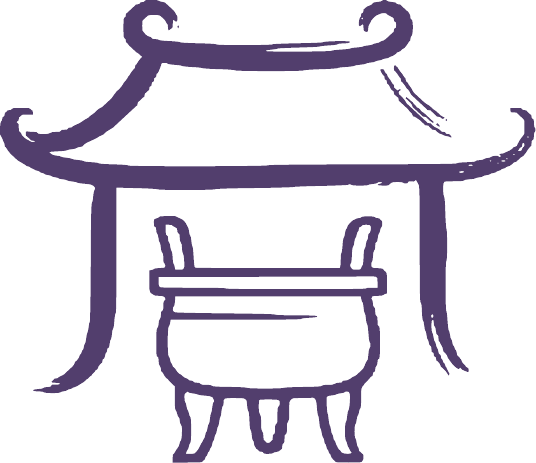}{scopeArtefact}{#4}%
}

\NewDocumentCommand{\entityTagText}{O{1.4em} O{-0.80ex} O{0.18em} m}{%
  \entityIconText[#1][#2][#3]{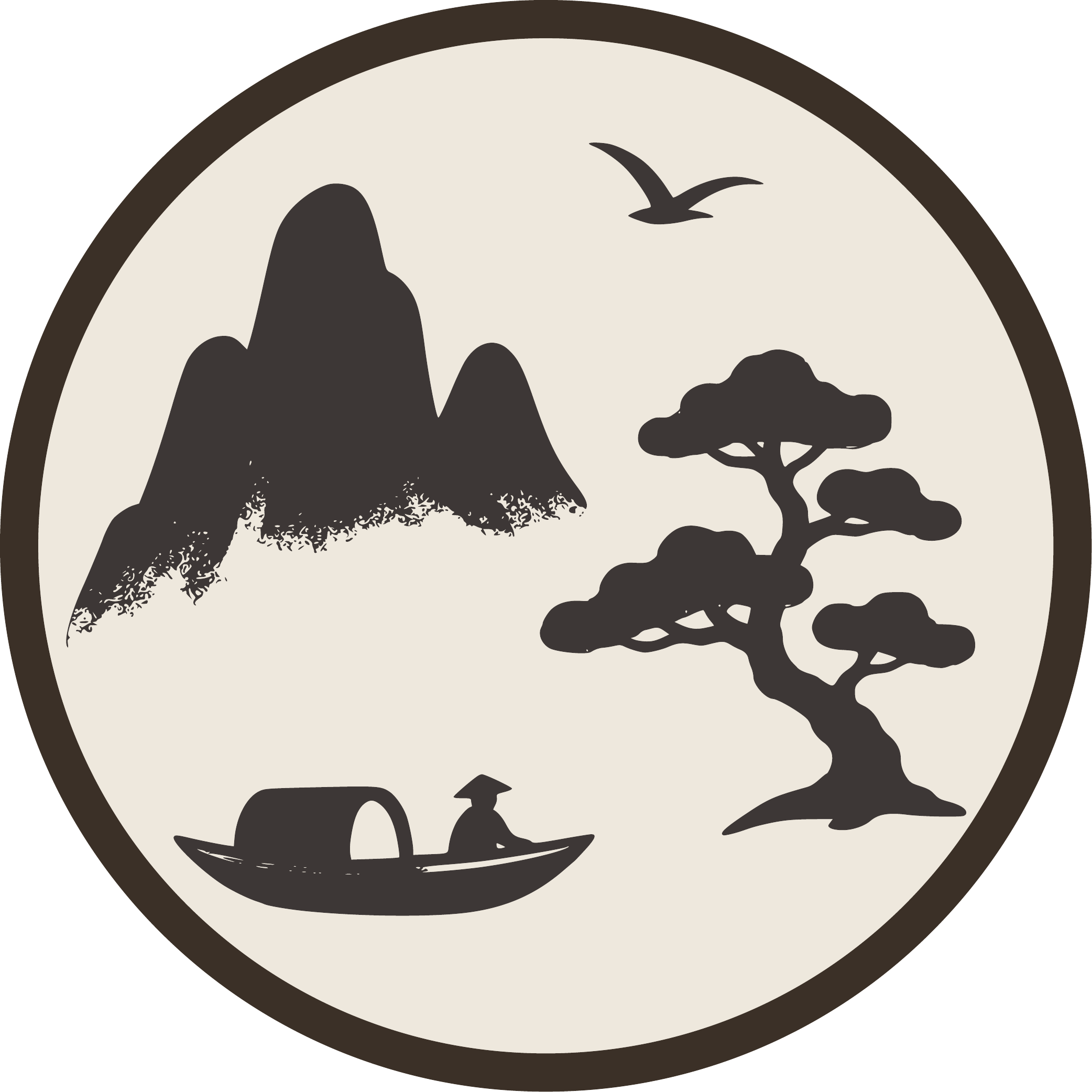}{scopeEntity}{#4}%
}

\NewDocumentCommand{\contextTagText}{O{1.4em} O{-0.80ex} O{0.18em} m}{%
  \entityIconText[#1][#2][#3]{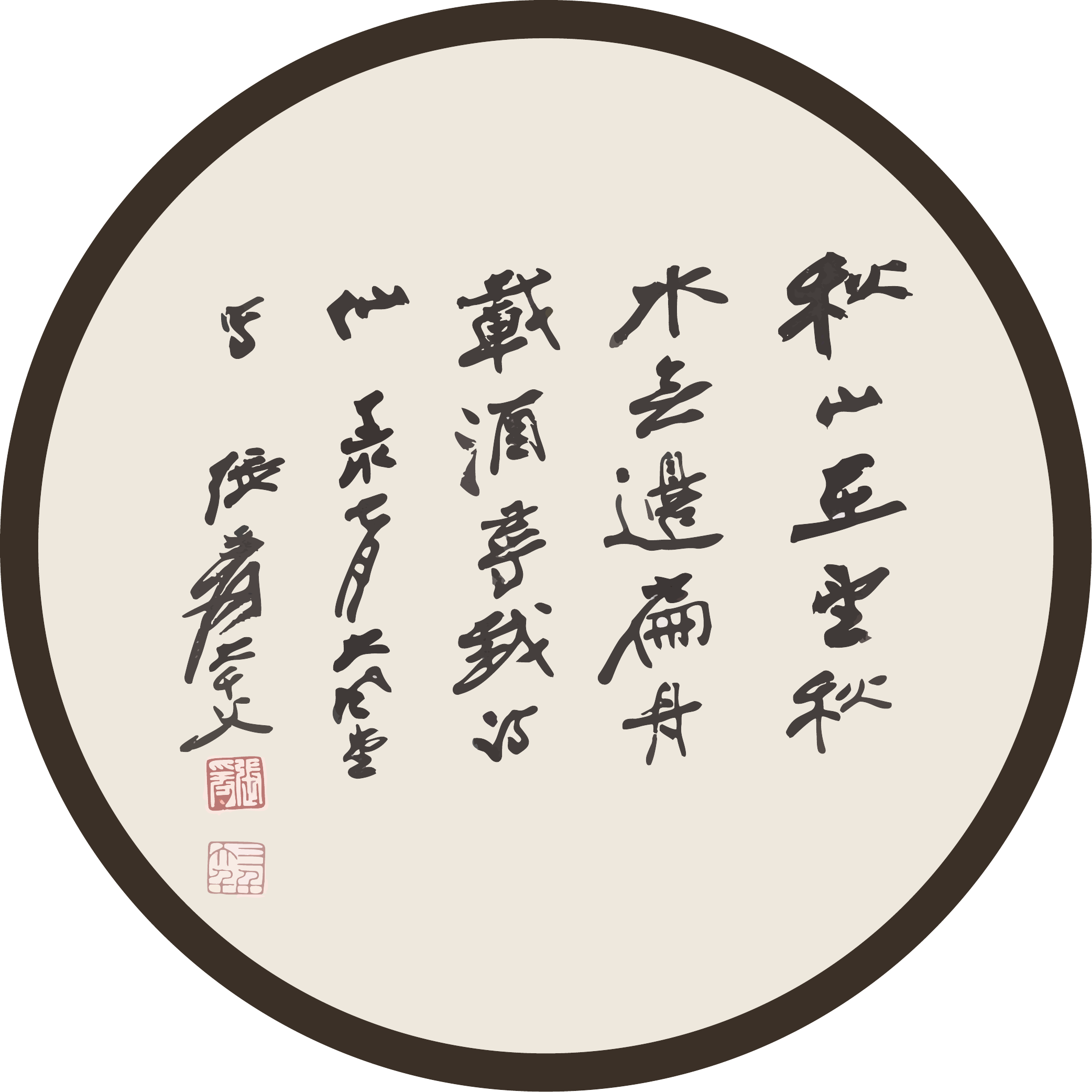}{scopeContext}{#4}%
}

\NewDocumentCommand{\voidTagText}{O{1.4em} O{-0.80ex} O{0.18em} m}{%
  \entityIconText[#1][#2][#3]{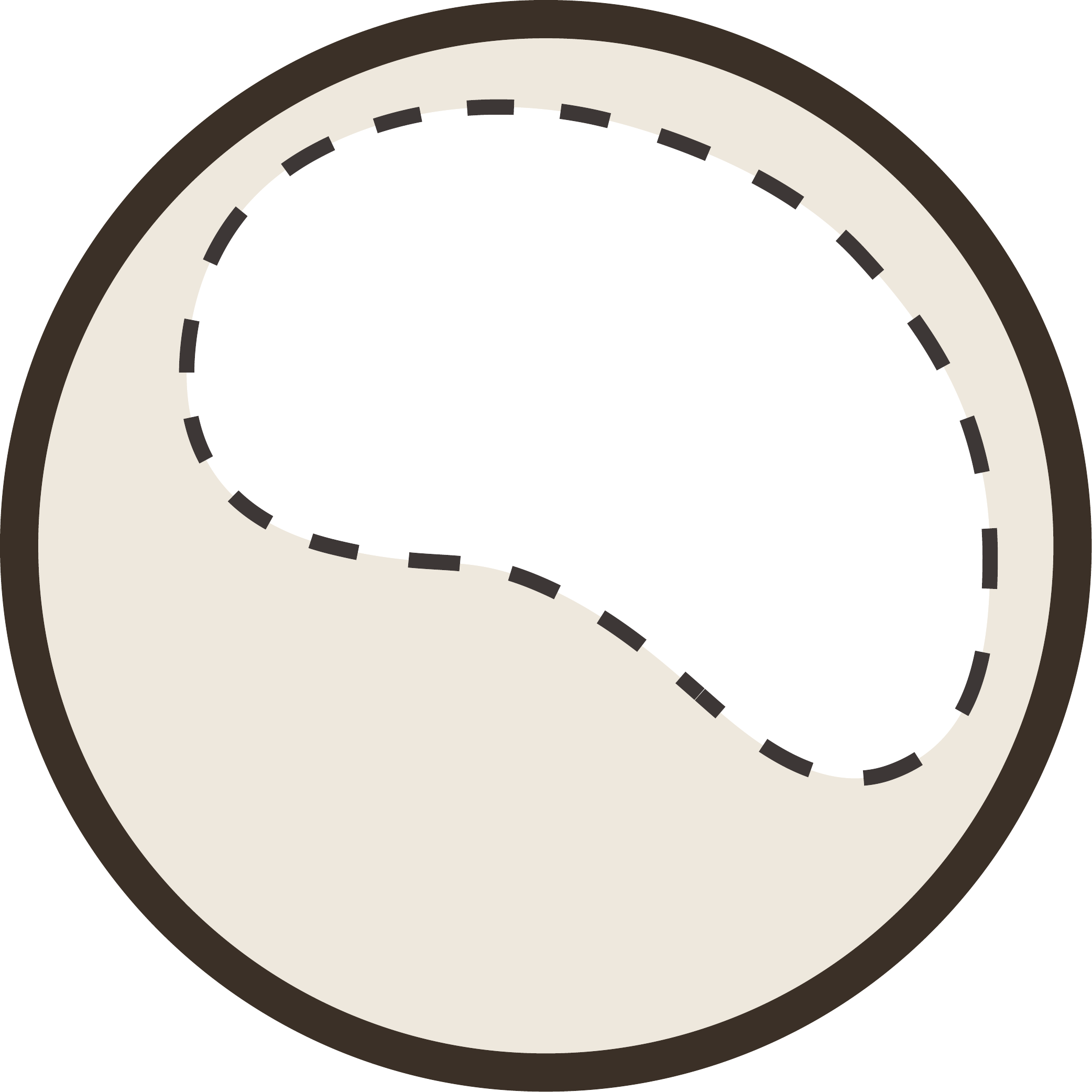}{scopeVoid}{#4}%
}

\NewDocumentCommand{\relationTagText}{O{1.4em} O{-0.80ex} O{0.18em} m}{%
  \entityIconText[#1][#2][#3]{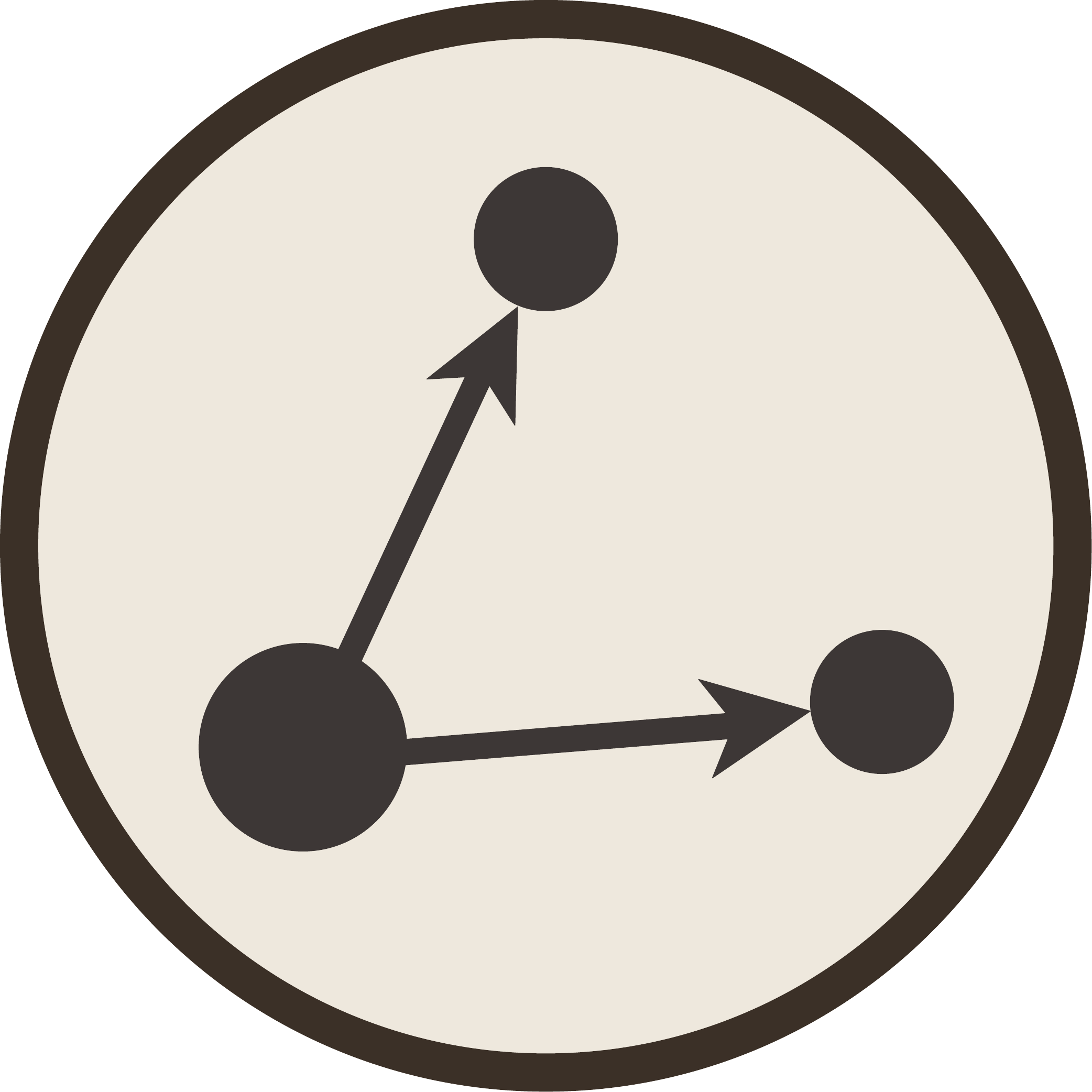}{scopeRelation}{#4}%
}

\newif\ifyrqreview
\yrqreviewtrue 
\ifyrqreview
  
  \newcommand{\yrqDelete}[1]{\textcolor{brown}{\sout{#1}}}
  
  \newcommand{\yrqComment}[1]{\textcolor[HTML]{0000CC}{[yrq: \begin{CJK*}{UTF8}{gbsn}#1\end{CJK*}]}}
\else
  
  \newcommand{\yrqDelete}[1]{}
  
  \newcommand{\yrqComment}[1]{}
\fi

\begin{document}

\title{\toolname: A Composition-Graph-Based Visual Analytics System for Compositional Analysis of Traditional Chinese Paintings}


\author{%
Dekun~Qian,
Ruiqi~Yu,
Li Ye,
Yize Li, 
Fengling~Zheng$^{\dagger}$,
Weigui Zheng, 

Yigang~Wang, Jinchang Li and Zhiguang~Zhou$^{\dagger}$%

\thanks{Dekun Qian and Ruiqi Yu contributed equally to this work.}
\thanks{Dekun Qian, Ruiqi Yu, Fengling Zheng, Yize Li, 
Yigang Wang, and Zhiguang Zhou are with Hangzhou Dianzi University, Hangzhou, China. E-mail: \{qiandekun, richyu, liyize, fenglingzheng,  
yigang.wang, zhgzhou\}@hdu.edu.cn.}
\thanks{Li Ye is with Zhejiang University, Hangzhou, China. E-mail: li-ye@zju.edu.cn.}
\thanks{Weigui Zheng is with Zhejiang Intangible Cultural Heritage Museum, Hangzhou, China. E-mail: zgsdbwg2009@163.com.}
\thanks{Jinchang Li is with Zhejiang University of Finance \& Economics, Hangzhou, China. E-mail: ljc@zufe.edu.cn.}
\thanks{$^{\dagger}$Corresponding authors: Fengling Zheng and Zhiguang Zhou.}%
}

\markboth{IEEE Transactions on Visualization and Computer Graphics}%
{Qian \MakeLowercase{\textit{et al.}}: \toolname{}}


\maketitle

\begin{abstract}
Compositional analysis of Traditional Chinese Paintings (TCPs) reveals
how spatial arrangement, narrative structure, and cultural-aesthetic
meaning are organized within the pictorial field. Traditional compositional analysis relies primarily on qualitative interpretation, supporting close examination of individual paintings but offering limited capacity to identify, compare, and validate compositional patterns across large-scale collections. To identify the key challenges in analyzing composition across large \TCP{} collections, we collaborated with two art historians and conducted a complementary literature review. Drawing on the resulting insights, we introduce the \compograph{}, a structured representation for composition-oriented analysis of \TCP{}s. It represents a painting's composition across four layers: entities, relations, voids, and context. Based on this representation, we develop \toolname{}, a canvas-based visual analytics system for composition-oriented exploration of \TCP{}s. \toolname{} allows art historians to construct and refine painting cohorts through interactive compositional queries. It also supports inspecting entity distributions and relations at the cohort level, comparing compositional differences across cohorts, and tracing aggregate patterns back to painting-level evidence.
Through two case studies, a user study, and expert
interviews, we demonstrate that \toolname{} can help art historians
discover, compare, and validate compositional patterns across
collections of \TCP{}s.
\end{abstract}


\section{Introduction}
\label{sec:introduction}

\begin{figure}[t]
    \centering
    \includegraphics[width=\columnwidth]{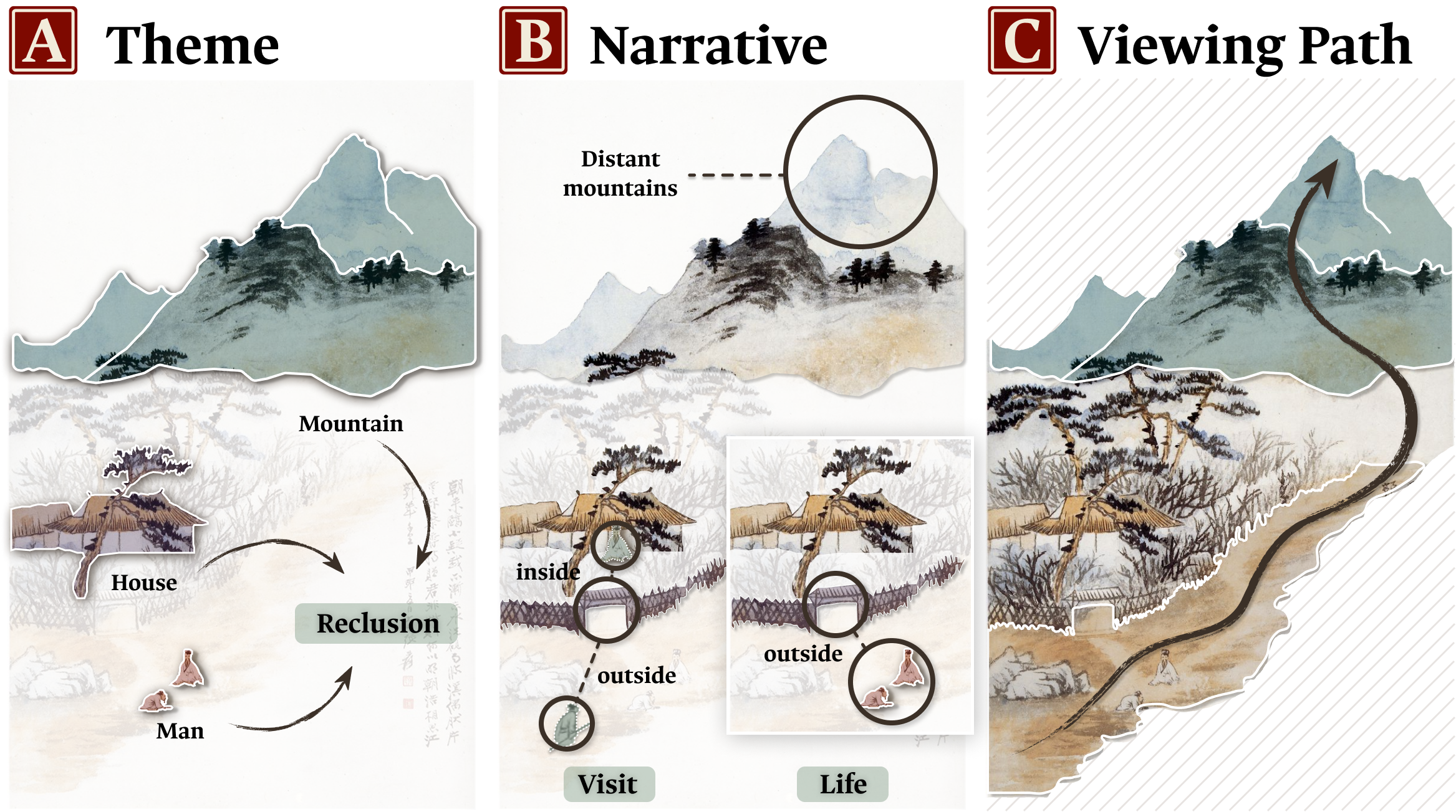}
    \caption{
    Composition supports three levels of art-historical reading.
    \paneltag{A}~Theme: mountains, dwelling, and figures suggest reclusion.
    \paneltag{B}~Narrative: relations among motifs determine how the same entities are interpreted as different narrative scenes, such as visiting or everyday life.
    \paneltag{C}~Viewing path: foreground, mountains, and void guide attention upward.
    }
    \label{fig:CompoA}
\end{figure}

\IEEEPARstart{C}{ompositional} analysis has long been central to art history, art theory, and painting practice, as it reveals how a work is structured, perceived, and interpreted. In Traditional Chinese Paintings (TCPs), the importance of composition is deeply rooted in Chinese art theory. Fifteen centuries ago, \textit{Xie He} identified composition and spatial arrangement as
one of the \textit{Six Principles of Painting}~\cite{Zhang_2024,Munakata_1968}. As illustrated in Fig.~\ref{fig:CompoA}, composition in TCP concerns how pictorial elements, such as figures, mountains, trees, and water, are arranged through spatial and semantic relations and balanced with voids to guide viewing paths and convey subject matter~\cite{arnheim1954art,liu2019qian,murray2007mirror}. Building on this theoretical tradition, art historians analyze composition through close reading of individual paintings and comparison across works by the same artist, works sharing the same subject matter, or works employing similar formats. Such analyses help identify layout tendencies, stylistic conventions, and recurring compositional patterns~\cite{jiang2012composition,Xue_2022}.

Traditional compositional analysis in \TCP{} relies primarily on qualitative interpretation through close reading of individual paintings and small groups of related works~\cite{silbergeld1982chinese,fong1992beyond}. While this approach enables nuanced art-historical interpretation, it provides limited support for systematically identifying and comparing recurring compositional patterns across large collections. The growing availability of digitized painting collections has motivated computational approaches for large-scale art analysis. Existing work has explored recurring visual arrangements through visual-link analysis and spatial matching~\cite{seguin2016visual,shen2019discovering}, compositional transfer through pose-based retrieval~\cite{jenicek2019linking}, and explicit modeling of image composition using action regions, action lines, and foreground--background organization~\cite{madhu2020understanding,madhu2023icc++}. Computational studies of \TCP{} have further advanced visual retrieval~\cite{Zhang_2024}, semantic annotation and knowledge organization~\cite{wan2024wumkg}, and multimodal painting understanding~\cite{chen2025multi}. Generic visual representations, such as scene graphs, also provide structured descriptions of depicted objects and their spatial or semantic relations~\cite{johnson2015image,chang2021comprehensive}. However, these approaches primarily focus on visual content or semantics rather than compositional organization, and therefore provide only limited support for composition-oriented analysis of \TCP{} collections.

To identify the representational and analytical requirements of composition-oriented \TCP{} analysis,
we conducted a formative study that combined a literature review with multiple rounds of semi-structured interviews with two art historians specializing in \TCP{}. Drawing on these sources, we identified two challenges (C1--C2) in supporting composition-oriented analysis of \TCP{}s at collection scale. 
\textbf{C1. Compositional evidence is difficult to formalize computationally.} 
Art historians interpret composition through the positions of pictorial elements, their spatial and semantic relations and the organization of voids within the pictorial field~\cite{jiang2012composition,silbergeld1982chinese,fong1992beyond}. Manual close reading can capture these structures but does not readily scale, while scalable proxies such as context or global image similarity discard the structure that makes the analysis compositional.
\textbf{C2. Compositional analysis requires iterative movement between close reading
and cross-painting comparison.} Compositional patterns
often become visible only through aggregation, yet can be validated only when traced back to the paintings behind them~\cite{arnold2019distant,manovich2020cultural,pirolli2005sensemaking}. The relevant painting cohorts and meaningful patterns emerge during analysis, rather than being fixed in advance. Cohort construction, comparison, and verification proceed together and are iteratively refined as the analysis develops.

To address these challenges, we introduce the \compograph{}, a structured representation that makes pictorial elements, spatial relations, voids, and context information explicit and queryable \textbf{(C1)}. Building on this representation, we develop \toolname{}, which supports an iterative workflow for forming cohorts, comparing aggregate patterns, and tracing them back to individual paintings for verification \textbf{(C2)}. Art historians can combine keyword-based search with interactively sketched compositional queries that specify entity positions, inter-entity relations, and voids to construct and refine painting cohorts. Coordinated views enable them to inspect entity distributions and
relations, compare compositional patterns across cohorts, and trace cohort-level findings back to individual paintings. We evaluated \toolname{} through two case studies, a user study, and expert interviews. The user study assessed workflow usability and perceived usefulness with domain participants; the case studies demonstrated how the system supports composition-oriented analysis in representative scenarios; and the expert interviews examined the domain validity and limitations of cohort construction, comparison, and painting-level evidence inspection. The results indicate that \toolname{} enables participants to construct and refine composition-oriented cohorts, identify and compare
compositional patterns, and trace cohort-level findings back to individual paintings.

The primary contributions of this work are summarized as follows:
\begin{itemize}
  \item We identify challenges and design requirements for composition-oriented \TCP{} analysis, and introduce the \compograph{} for compositional analysis.

  \item We develop \toolname{}, a canvas-based visual analytics system that integrates scene-graph-based composition modeling and retrieval with coordinated views of entity distributions, relations, and cohort differences.

  \item We evaluate \toolname{} through two case studies, a user study, and expert interviews, demonstrating its support for an integrated workflow that connects structured compositional
representation and querying with interactive compositional analysis.
\end{itemize}

\section{Related Work}
\label{sec:rw}

\subsection{Compositional Analysis in Traditional Chinese Painting}

Composition is a central concern in the study of \TCP{}s. Rather than examining depicted motifs in isolation, art historians interpret how pictorial elements are arranged within the pictorial field to create spatial organization, visual balance, narrative meaning, and artistic expression~\cite{silbergeld1982chinese}. This understanding is reflected in the classical concept of \textit{jingying weizhi} ("planning and placing"), which emphasizes the organization of pictorial elements rather than their mere depiction~\cite{Zhang_2024}. Chinese painting theory further characterizes composition through principles such as Guo Xi's \textit{three distances}, which describe pictorial space~\cite{bush2012early}, as well as the balance between solid and void, which treats empty space as an active compositional element~\cite{hearn2008read}. Other concepts, including opening--closing and host--guest relationships, further explain how visual emphasis is distributed throughout the picture~\cite{barnhart1997three,fong1992beyond}.

Composition is also inherently comparative. Rather than interpreting paintings individually, art historians examine works across artists, dynasties, subjects, and painting formats to identify recurring compositional strategies and stylistic tendencies~\cite{cahill1994painter}. Such comparisons focus on patterns in motif placement, spatial balance, visual pathways, and the use of void~\cite{hearn2008read}. Researchers also interpret these patterns within their historical and artistic contexts to understand how different schools and artists developed distinctive visual organizations~\cite{fong1992beyond}. However, this process remains largely qualitative and manual. Scholars repeatedly move between close inspection of individual paintings and comparison across evolving painting groups, making composition-oriented analysis increasingly difficult to scale as digitized collections continue to grow.

\subsection{Computational Approaches for Compositional Analysis}
\label{sec:rw_computational_composition}

Computational approaches have increasingly explored pictorial composition through spatial and structural representations rather than global image appearance. Early work employed spatially consistent matching to discover recurring visual arrangements across artworks~\cite{furuta2019efficient}. Pose-based retrieval enabled the analysis of compositional transfer between paintings by matching human configurations across collections~\cite{jenicek2019linking}. More recently, Image Composition Canvas modeled action regions, action lines, and foreground--background organization to support art-historical image retrieval and interpretation~\cite{madhu2020understanding,madhu2023icc++}. Complementing these approaches, scene graphs represent pictorial entities and their semantic or spatial relations~\cite{johnson2015image}, while graph matching supports retrieval according to structural correspondence rather than visual similarity~\cite{yoon2021image}. Together, these studies demonstrate the value of representing compositional structure explicitly for large-scale artwork analysis.

For \TCP{}s, computational research has mainly focused on visual retrieval~\cite{jiang2021mtffnet}, brushstroke and style analysis~\cite{sun2015brushstroke}, knowledge graph construction~\cite{wan2024wumkg}, and multimodal understanding of paintings~\cite{feng2022ipoet}. Although these methods improve access to large painting collections, they primarily emphasize depicted content, artistic style, or contextual knowledge instead of composition itself. Existing composition-oriented representations, meanwhile, have largely been developed for general artworks rather than \TCP{}s. Consequently, computational support for representing, querying, and comparing compositional structures in \TCP{} collections remains limited.

\subsection{Visual Analytics for Cultural Heritage}
\label{sec:rw_va}

Visual analytics has become an important approach for exploring cultural heritage collections and supporting scholarly interpretation~\cite{windhager2018visualization}. Collection-oriented systems such as VIKUS enable interactive exploration of museum collections through coordinated visual layouts~\cite{glinka2017past}. SeMap supports semantic exploration of cultural collections by organizing metadata and contextual relationships~\cite{portales2022increasing}, while MuseKG facilitates knowledge-driven exploration through cultural heritage knowledge graphs~\cite{li2025musekg}. Beyond collection exploration, recent systems have investigated artwork relationships~\cite{zhang2024scrolltimes}, evolutionary analysis of artistic styles~\cite{oda2025artevoviewer}, and detailed inspection of visual characteristics within individual paintings~\cite{fan2020visual,sun2015brushstroke,feng2022ipoet}. Together, these studies demonstrate the value of coordinated visual representations for cultural heritage analysis.

Despite this progress, these research directions remain largely disconnected. Image Composition Canvas explicitly models pictorial composition for artwork retrieval~\cite{madhu2020understanding,madhu2023icc++}, but places less emphasis on iterative cohort analysis. More generally, visual analytics research has shown that interactive querying~\cite{stolte2002} and iterative refinement~\cite{wongsuphasawat2015voyager} are essential for exploratory analysis and sensemaking. Taken together, existing work provides limited support for integrated composition-oriented analytical workflows across large painting collections.

\section{Background}
\label{sec:designreq}


This section presents the findings of our formative study. Drawing on a six-month collaboration with two art historians and a structured literature review, we first identify the analytical dimensions underlying composition-oriented analysis of \TCP{}s. Based on these findings, we derive two key challenges (C1--C2), which inform two design requirements (R1--R2) and six analytical tasks (T1--T6) that guide the system design.

\subsection{Formative Study}
\label{sec:bg_method}

To inform the design of \toolname{}, we conducted a formative study that combined a six-month collaboration with two domain experts and a structured literature synthesis.

\subsubsection{Collaboration with Art Historians}
We collaborated for about 6 months with two art historians (\textbf{A1} and \textbf{A2}) who specialize in \TCP. \textbf{A1} is a university professor with 23 years of experience in Chinese art history and traditional painting studies, while \textbf{A2} is a museum professional with over 20 years of experience in Chinese painting collections and curatorial practice. Their expertise covers both scholarly research and collection-facing interpretation. Initially, we conducted in-depth semi-structured interviews and open discussions about their workflows, analytical practices, research barriers, and expectations of computational support. 
During the later stage, we worked with the experts to refine the design requirements and analytical tasks through iterative discussions on emerging design concepts, query representation, cohort construction, and visual analysis strategies. All sessions were audio-recorded with consent and transcribed. Two researchers independently coded the transcripts using thematic analysis~\cite{braun2006using} and reached consensus on a shared codebook.

\subsubsection{Literature Review}

To complement our collaboration with the art historians, we conducted a structured literature review~\cite{snyder2019literature,okoli2015guide}. The review covered work published between 1990 and 2025 across major visualization, computer vision, digital humanities, and art history venues, complemented by Google Scholar searches~\cite{haddaway2015role}. We used combinations of keywords related to (i) \emph{traditional Chinese painting and composition} (e.g., ``traditional Chinese painting'', ``Chinese painting'', ``composition'', ``art history''), (ii) \emph{computational representations and analysis} (e.g., ``scene graph'', ``knowledge graph'', ``painting retrieval'', ``image retrieval''), and (iii) \emph{visual and cultural analytics} (e.g., ``visual analytics'', ``cultural analytics'', ``digital art history''). These search and screening steps followed established guidance for structured literature reviews~\cite{kitchenham2007guidelines}. Beyond venue- and keyword-based retrieval, we further traced forward and backward citations from influential papers to improve coverage~\cite{wohlin2014guidelines}. We focused on studies involving \TCP{}s, compositional analysis, structured visual representations, computational image analysis, and interactive visual analysis. In total, 76 candidate papers were identified, of which 57 representative papers were retained after removing duplicates and excluding work not directly relevant to composition-oriented analysis. We followed established guidelines for conducting and structuring literature reviews~\cite{templier2015framework}, and organized the reviewed studies using a multidimensional coding scheme~\cite{Ye2025HyperMOOCAM}. We adapted this framework to the context of computational analysis of \TCP composition and coded the retained papers along four dimensions: research field, analytical task, employed data, and content of interest. Each paper could receive multiple labels within a dimension. The ``Mixed data'' label was assigned to studies involving multiple modalities and could co-occur with individual data types. The distribution of the retained papers across these dimensions is summarized in Tab.~\ref{tab:literature_review}. We combined the review findings with insights from our collaboration with art historians to identify the challenges that guided the subsequent design process.


\begin{table}[tb]
\centering
\caption{Distribution of coded categories for the 57 reviewed papers.}
\label{tab:literature_review}
\setlength{\tabcolsep}{4pt}
\renewcommand{\arraystretch}{1.05}
\footnotesize

\begin{tabular}{p{0.35\linewidth}@{\hspace{2pt}}c
                p{0.35\linewidth}@{\hspace{2pt}}c}

\toprule

\multicolumn{2}{c}{\textbf{Research Field}}
&
\multicolumn{2}{c}{\textbf{Analytical Task}}
\\

Category & Count & Category & Count\\

\midrule

Art history                  & 8  & Interpretation   & 13\\
Computational analysis       & 15 & Representation   & 31\\
Structured representation    & 11 & Retrieval        & 23\\
Digital art history          & 4  & Exploration      & 28\\
Cultural analytics           & 9  & Comparison       & 31\\
Analysis workflow            & 10 & Verification     & 21\\

\midrule

\multicolumn{2}{c}{\textbf{Employed Data}}
&
\multicolumn{2}{c}{\textbf{Content of Interest}}
\\

Category & Count & Category & Count\\

\midrule

TCP images               & 17 & Entities               & 39\\
Artwork images           & 18 & Spatial organization   & 24\\
Scene graphs             & 9  & Relations              & 31\\
Metadata                 & 38 & Void                   & 9\\
Interaction data         & 15 & Context                & 30\\
Mixed data               & 13 & Composition patterns   & 40\\

\bottomrule

\end{tabular}
\end{table}

\subsection{Findings from the Formative Study}
\label{sec:findings}

The interviews and literature review first helped us identify the
analytical dimensions that art historians consider when interpreting
composition in \TCP{}s. Building on this shared
understanding, we then identified two recurring challenges that limit
computational support for composition-oriented analysis.

\subsubsection{Analytical Dimensions of Composition}
\label{sec:bg_composition}


Both experts described compositional analysis as understanding how
pictorial elements are organized to create spatial structure, visual
balance, and artistic meaning. This view is consistent with \TCP theory, which emphasizes motif arrangement and the
organization of pictorial space as the foundation of composition
~\cite{fong1992beyond,silbergeld1982chinese}. Art-historical studies further highlight the interplay between painted
forms and void, as well as the importance of historical and material
context in interpreting compositional choices
~\cite{hearn2008read,murck1991words,delbanco2000chinese}.


Based on the literature review, we first identified seven candidate analytical dimensions frequently discussed in composition research: pictorial motifs, spatial organization, narrative relations, void, viewing format, artistic style, and historical context. We then used these dimensions to guide a series of semi-structured interviews with the two art historians. Following iterative thematic coding~\cite{Braun01012006}, we analyzed which dimensions were consistently referenced during experts' composition-oriented reasoning and could be represented computationally. Narrative relations were integrated into entity relationships, while viewing format, artistic style, and historical context were consolidated as painting-level contextual information. This process resulted in four computational dimensions: Entities, Relations, Voids, and Context, which form the basis of the subsequent representation.

\textit{Pictorial Motifs.}
Art historians first examine the pictorial motifs that constitute a
painting, such as figures, mountains, trees, architecture, boats, and
animals. Attention is paid not only to what motifs appear but also to
how they are selected, scaled, and positioned within the pictorial
field, revealing compositional balance and recurring layout patterns
across paintings
~\cite{fong1992beyond,silbergeld1982chinese,hearn2008read}.

\textit{Spatial Organization.}
Composition also depends on the spatial organization of motifs. Relative position, overlap, distance, orientation, and interactions establish visual hierarchy, narrative structure, and viewing flow~\cite{fong1992beyond,jiang2012composition}.

\textit{Solid--Void Organization.}
Chinese painting places particular emphasis on the dynamic balance
between painted forms and reserved blank space. Rather than functioning
as empty background, void contributes to visual rhythm, spatial depth,
and expressive balance, while different compositional strategies vary in
how extensively it is employed
~\cite{hearn2008read,murck1991words,fan2019,xiao2026square}.

\textit{Art-historical Context.}
Compositional choices are interpreted within broader historical and
material contexts, including artist, period, subject, painting format,
material, and collection information
~\cite{murck1991words,delbanco2000chinese,fong1992beyond}. Such context
determines which paintings constitute meaningful grounds for comparison
and explains how compositional tendencies vary across artistic and
historical settings.

\subsubsection{Challenges for Composition-Oriented Analysis}

Although these four dimensions collectively define how composition is interpreted in art history, our interviews and literature review reveal that current computational approaches provide only limited support for representing and analyzing them together. Two recurring challenges emerged.

\textbf{C1. Compositional evidence is difficult to formalize computationally.}
Current computational studies of \TCP{}s mainly focus on retrieval, classification, captioning, style analysis, and metadata organization, while computational support for representing compositional structure remains limited~\cite{dong2020,wan2024wumkg}. Although relation-aware representations such as scene graphs encode objects and their relationships, they are not designed to capture the broader compositional evidence used in art-historical analysis, including spatial organization, visual balance, and the role of reserved space~\cite{johnson2015image,krishna2017visual}. As \textbf{A2} explained, ``\textit{Current databases record who painted a work and what it contains, but they rarely describe how the composition is organized.}'' Consequently, composition remains difficult to represent as structured, queryable visual evidence, limiting composition-oriented retrieval and analysis.

\vspace{0.4em}

\textbf{C2. Compositional analysis requires iterative movement between close
reading and cross-painting comparison.}
Both experts described compositional analysis as an iterative process that alternates between forming hypotheses, constructing working collections, comparing multiple cohorts, and returning to individual paintings for verification. As \textbf{A1} explained, ``\textit{It is difficult to understand a compositional style from a single painting. We usually compare many paintings before drawing a conclusion.}'' Similarly, \textbf{A2} noted, ``\textit{After finding several relevant paintings, I often need to go back and adjust my search to refine the collection.}'' These observations are consistent with research on cultural analytics and sensemaking, which characterizes visual analysis as an iterative process alternating between distant viewing, comparison, and close inspection~\cite{arnold2019distant,manovich2020cultural,pirolli2005sensemaking}. However, existing computational tools provide limited support for this iterative analytical workflow.

\subsection{Requirements and Task Analysis}
\label{sec:bg_dr}


Based on these challenges, we formulated two design requirements and iteratively refined them with the experts. We further translated them into six analytical tasks.


\noindent
\textbf{R1.} \textbf{Represent compositional evidence as structured and inspectable data.} To address C1, the system should translate the compositional dimensions identified in the formative study into structured, inspectable representations that remain available throughout querying, cohort summarization, and cross-cohort comparison, enabling scholars to verify how analytical results are derived.


\hangindent=2.5em\hangafter=1
\noindent\hspace*{0.8em}
\textbf{T1.}~\textbf{Interpret the compositional structure of individual paintings.}
Support inspection of how entities, spatial arrangements, void, and contextual information jointly form the composition of a painting.

\hangindent=2.5em\hangafter=1
\noindent\hspace*{0.8em}
\textbf{T2.}~\textbf{Summarize compositional distributions and relationships within a cohort.}
Help art historians examine where entities tend to appear, which entities frequently occur together, and what relations organize a cohort.

\hangindent=2.5em\hangafter=1
\noindent\hspace*{0.8em}
\textbf{T3.}~\textbf{Compare compositional tendencies across comparable cohorts.}
Help art historians compare how spatial distributions and relational structures differ across cohorts defined by artist, period, subject, or format.

\noindent
\textbf{R2.}~\textbf{Support iterative comparison and evidence verification.} To address C2, the system should support iterative exploration rather than a fixed query-and-result workflow. It should enable art historians to progressively formulate and refine compositional hypotheses, construct and revise comparison cohorts, and repeatedly move between cohort-level summaries and painting-level evidence during analysis.

\hangindent=2.5em\hangafter=1
\noindent\hspace*{0.8em}
\textbf{T4.}~\textbf{Query and refine compositional structures.}
Help art historians express hypotheses through entities, positions, relations, voids, and context filters, and refine these structures during exploration.

\hangindent=2.5em\hangafter=1
\noindent\hspace*{0.8em}
\textbf{T5.}~\textbf{Construct and revise composition-oriented cohorts.}
Support the creation, filtering, splitting, and revision of cohorts as art historians move across analytical categories and refine their scope.

\hangindent=2.5em\hangafter=1
\noindent\hspace*{0.8em}
\textbf{T6.}~\textbf{Verify cohort-level patterns with painting-level evidence.}
Enable art historians to return from cohort-level summaries and comparisons to paintings, matched entities, and supporting relations for verification.

\section{Composition Modeling and Analysis}
\label{sec:method}

We transform digitized \TCP{} images, annotations, and context into a \compograph{} that supports an iterative workflow of compositional analysis (Fig.~\ref{fig:pipeline}).

\subsection{\compograph{}}
\label{sec:composition_graph}

Informed by R1 and R2, we introduce the \compograph{}, a structured representation of the composition of a painting. Building on the analytical dimensions identified in Sec.~\ref{sec:bg_composition}, we represent pictorial motifs as \entityTagText{Entities}, spatial organization through entity locations and \relationTagText{Relations}, solid--void organization as \voidTagText{Voids}, and art-historical context as \contextTagText{Context}. Together, these four layers form a unified computational representation for composition-oriented analysis.
Since entities and their relations naturally form a graph structure, we adopt the scene graph~\cite{johnson2015image,krishna2017visual} as the backbone of our representation. We further extend it with two composition-specific layers that explicitly represent void and art-historical context, resulting in the \compograph{} (T1).

\subsubsection{Four-Layer Representation}
\begin{figure}[t]
    \centering
    \includegraphics[width=0.96\columnwidth]{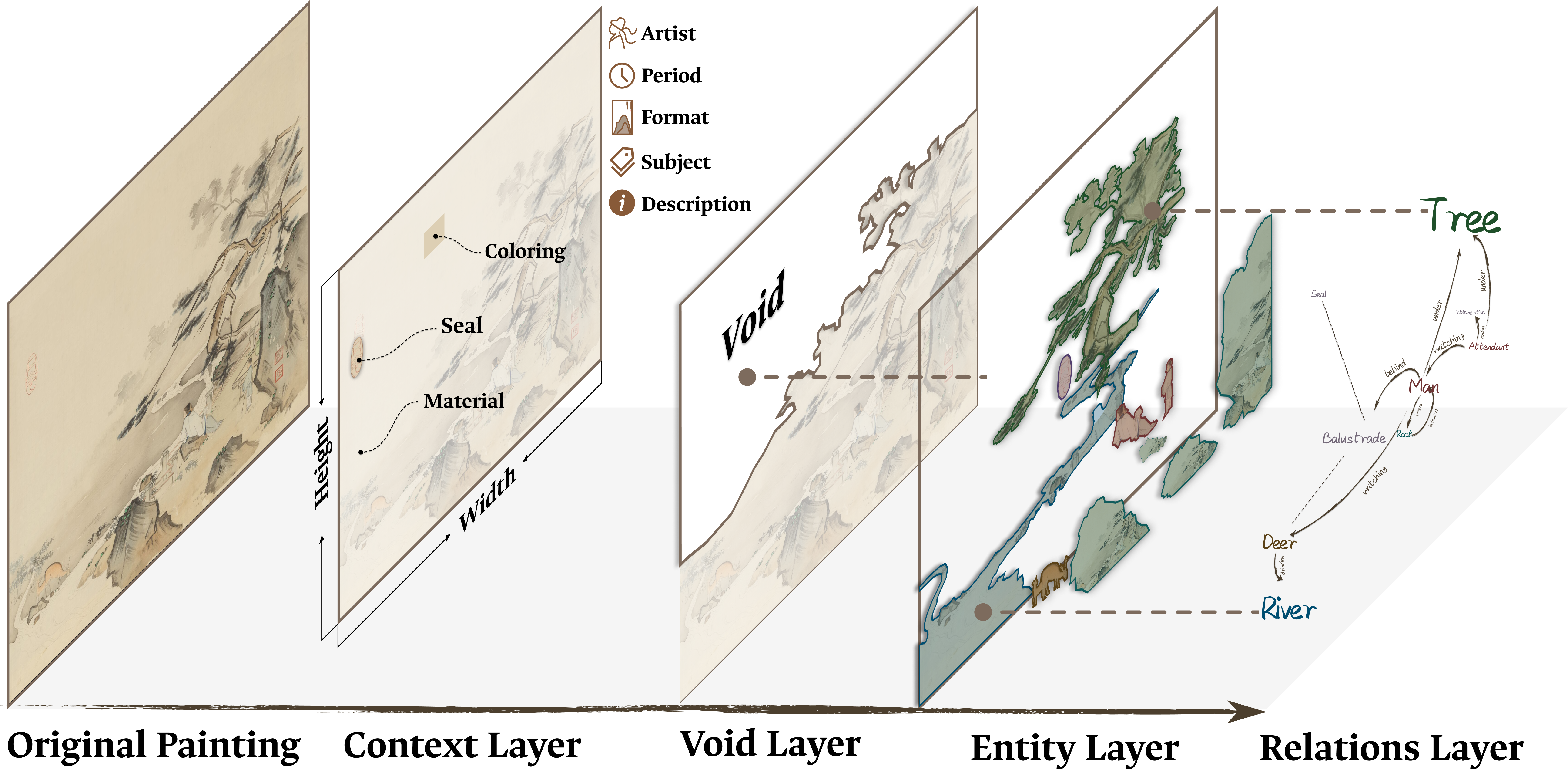}
    \caption{
       \compograph{} of a modern copy after \textit{Ma Yuan}'s \textit{Songyin Guanlu Tu}. The graph consists of four layers: the entity layer records elements such as the seated figure, pine, rocks, bridge, and water; the relation layer encodes their spatial and semantic relations; the void layer records the broad upper-left void; and the context layer stores painting-level information such as the title, artist, period, format, material and seal.
    }
    \label{fig:compositiongraph}
\end{figure}

For each painting $p$, the \compograph{} is
\begingroup
\setlength{\abovedisplayskip}{3pt}
\setlength{\belowdisplayskip}{3pt}
\begin{equation}
G_p=(E_p,R_p,V_p,C_p),
\end{equation}
\endgroup
where $E_p$ is the entity layer, $R_p$ the relation layer, $V_p$ the void layer, and $C_p$ the context layer. The entity and relation layers follow the basic object relation structure of image scene graphs, where entities are represented as nodes and their spatial or semantic relations are represented as edges~\cite{johnson2015image,krishna2017visual,chang2021comprehensive}. We adapt this backbone to \TCP{} composition by using a \TCP{}-specific entity vocabulary and relation types derived from semantic annotation of \TCP{}s~\cite{vistcp,feng2022ipoet}. The void and context layers extend the scene-graph structure to capture composition-specific evidence that is central to \TCP{} analysis but is not represented in generic scene graphs~\cite{Zhang_2024,silbergeld1982chinese,fong1992beyond}. Fig.~\ref{fig:compositiongraph} shows an example annotated painting.

\noindent\textbf{\entityTagText{Entity layer}.}
The entity layer $E_p$ models pictorial elements as graph nodes. We build on the four-class semantic annotation scheme of VisTCP~\cite{vistcp}. Through iterative discussions with collaborating art historians and consideration of the compositional structures represented in our corpus, we refine its broad \textit{nature scenery} category into three categories: \entityVegetationTagText{vegetation}, \entityLandformTagText{landform}, and \entityWaterTagText{water}. Together with \entityHumanTagText{human figures}, \entityAnimalTagText{animals}, and \entityArtefactTagText{artefacts}, the resulting taxonomy contains six entity categories tailored to composition-oriented analysis. Each node stores its semantic category together with a bounding region and normalized position and size within the painting. These nodes provide the spatial primitives for subsequent retrieval, summarization, and comparison.


\noindent\textbf{\relationTagText{Relation layer}.}
The relation layer $R_p$ models compositional structure as typed edges between entity nodes. Following VisTCP~\cite{vistcp}, we distinguish two categories of relations: location relations describing spatial arrangements (e.g., above, behind, along, inside) and event relations describing semantic interactions (e.g., holding, sitting on, watching). Representing relations explicitly enables compositions to be retrieved and compared according to structural organization rather than entity presence alone.


\noindent\textbf{\voidTagText{Void layer}.}
The void layer $V_p$ represents reserved regions as independent spatial primitives. Each void is stored in the same normalized coordinate system as entity nodes, allowing painted and unpainted regions to be analyzed within a unified spatial representation. This layer makes void space an explicit and queryable component of the compositional representation, enabling its use as spatial evidence in subsequent analysis.


\noindent\textbf{\contextTagText{Context layer}.}
The context layer $C_p$ stores painting-level metadata as attribute-value pairs associated with each \compograph{}. It includes title, artist, period, subject, format, material, inscription, seal, collection, and related description~\cite{fong1992beyond,zhang2024scrolltimes}. Unlike the spatial layers, the context layer provides non-geometric evidence for filtering, cohort construction, and composition-oriented comparison by connecting visual patterns with their art-historical setting.

\begin{figure}[t]
    \centering
    \includegraphics[width=\columnwidth]{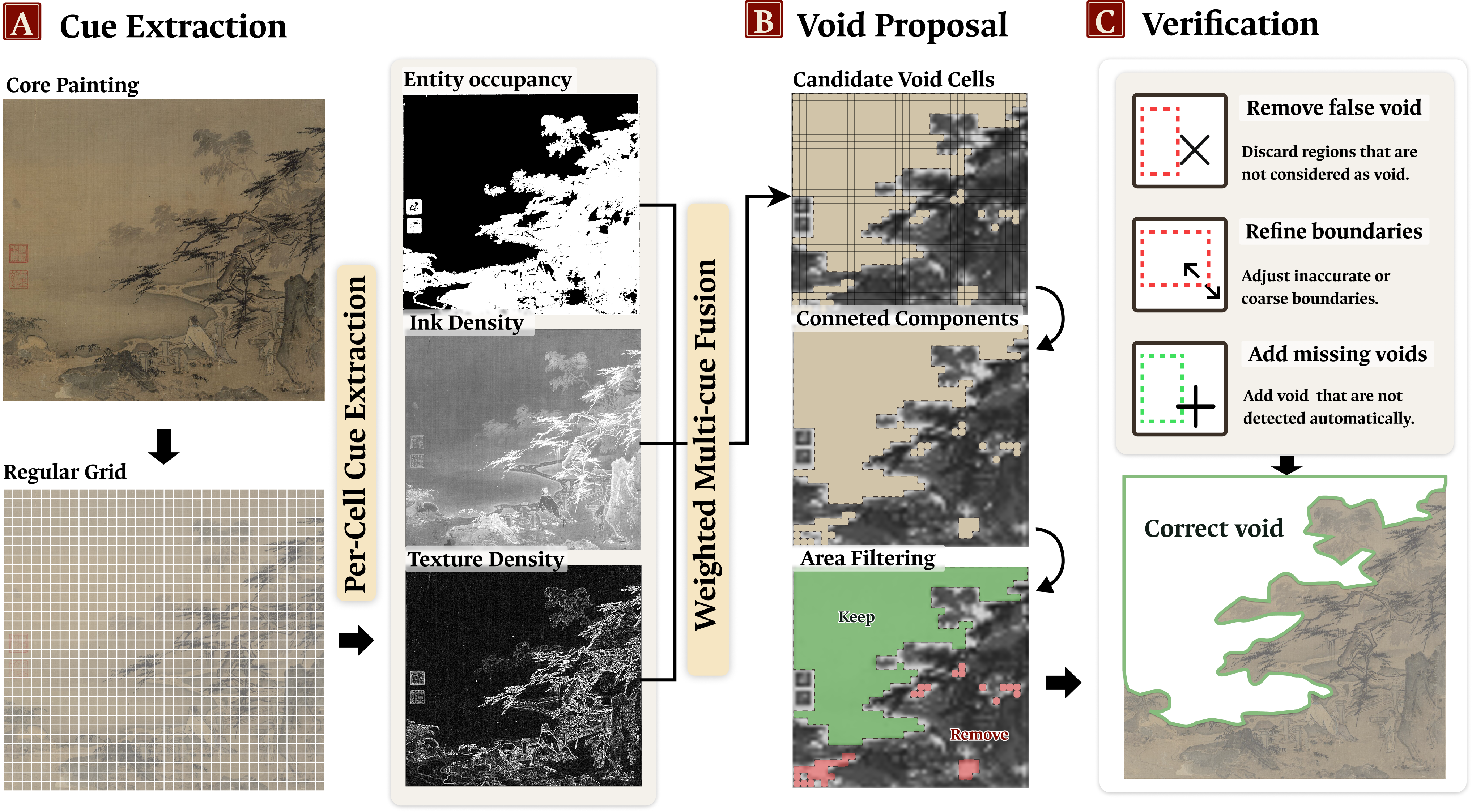}
    \caption{
    Construction of the void descriptor.
    \paneltag{A}~Three cues are extracted from grid cells.
    \paneltag{B}~Cue fusion and region filtering produce candidate voids.
    \paneltag{C}~Annotators verify and refine the candidates.
    }
    \label{fig:voidspace}
\end{figure}

\subsubsection{Data and Annotation}
\label{sec:dataset}
We instantiate the \compograph{} on a curated \TCP{} corpus annotated following VisTCP~\cite{vistcp}. Our dataset was constructed from the Chinese Ancient Open Dataset (CAOD)~\cite{caod2021}, an open repository maintained by the Chinese Treasures Museum that provides curated metadata and high-resolution images of \TCP{}s. We focus on bounded formats such as hanging scrolls, album leaves, and round fans, whose composition can be analyzed within a defined canvas. Handscrolls are excluded because their sequential viewing structure differs fundamentally from these formats. We selected 960 paintings with complete metadata and annotated entities, relations, and voids. The resulting dataset contains 7,587 entity instances, 6,180 relation instances, and 589 annotated voids. The taxonomy comprises 109 fine-grained entity labels organized into six semantic classes and 37 relation types organized into event and location relations.

Annotation was conducted by three annotators, including two domain experts in \TCP and one co-author with extensive expertise in the history and practice of \TCP. The annotation process followed the four-layer structure of the \compograph{}. Entity and relation annotations were manually assigned using predefined taxonomies, and context fields were transcribed from the corresponding painting metadata. For the void layer, we adopted an image-based proposal-and-verification procedure (Fig.~\ref{fig:voidspace}) based on standard image processing techniques~\cite{gonzalez2009digital}. After manually determining the core painting area, the system divided the painting into a regular grid and computed three complementary cues for each grid cell: entity occupancy derived from annotated object regions, ink density estimated from normalized grayscale intensity, and texture density estimated using the Sobel gradient operator~\cite{sobel19683x3}. These cues were combined through a weighted multi-cue fusion strategy commonly used in image analysis~\cite{kong2015} to identify candidate voids. Candidate cells were then grouped using connected-component analysis~\cite{he2008run}, and small fragmented regions were removed through area filtering. The resulting regions were treated only as void proposals. The annotation team subsequently reviewed these proposals by removing false positives, refining inaccurate boundaries, and adding missing regions when necessary, producing the final void annotations.

To assess inter-annotator agreement, two domain experts independently re-annotated a subset of 120 paintings selected from the completed dataset. Categorical agreement for entity and relation labels was measured using Cohen's $\kappa$, while spatial agreement for entity bounding regions and voids was evaluated using intersection over union (IoU). The resulting agreement scores were $\kappa=0.86$ for categorical labels and an average IoU of 0.78 for spatial regions. For the proposal-based void annotation, 32\% of the initial void proposals required manual intervention, including removal of false positives or refinement of region boundaries, and the annotators identified 13 additional voids that were not generated by the initial proposal procedure. Disagreements between the two independent re-annotations were subsequently reviewed with the third annotator and resolved through consensus. Tab.~\ref{tab:dataset} summarizes the resulting dataset used to construct the \compograph{}.

\begin{table}[t] \centering \caption{Summary of the curated working dataset.} \label{tab:dataset} \footnotesize \setlength{\tabcolsep}{3pt} \renewcommand{\arraystretch}{1.12} \begin{tabularx}{\columnwidth}{@{}>{\raggedright\arraybackslash}p{0.28\columnwidth} >{\raggedright\arraybackslash}X@{}} \toprule Item & Statistics \\ \midrule Paintings & 960 \TCP{} paintings with complete metadata. \\ Historical coverage & Tang to modern periods; major groups include Southern Song (248), Qing (185), Ming (177), Yuan (109), and Northern Song (102). \\ Themes & Landscape (444), Figure (273), Flower-and-bird (137), Animals (27), and Other/Unclassified (79). \\ Formats & Hanging scroll (660), Album leaf (122), Round fan (41), Handscroll/long scroll (44), and Other or unspecified (93). \\ Media & Silk (583), Paper (357), and Other or unspecified (20). \\ Top subject labels & Tree (231), Landscape (228), Figure (193), Architecture (165), River/lake/sea (144), Boat (105), and Fisherman/boatman (102). \\ \bottomrule \end{tabularx} \end{table}

\subsection{\compograph{}-Based Analysis}
\label{sec:composition_analysis}

\subsubsection{Composition Query and Retrieval}
Experts often begin with a tentative compositional hypothesis and revise it as the analysis develops. We support this process with an editable query over the four \compograph{} layers. The retrieved and ranked paintings can then be organized into a reusable cohort for subsequent analysis (T4, T5).

\begin{figure*}[t]
    \centering
    \includegraphics[width=0.96\textwidth]{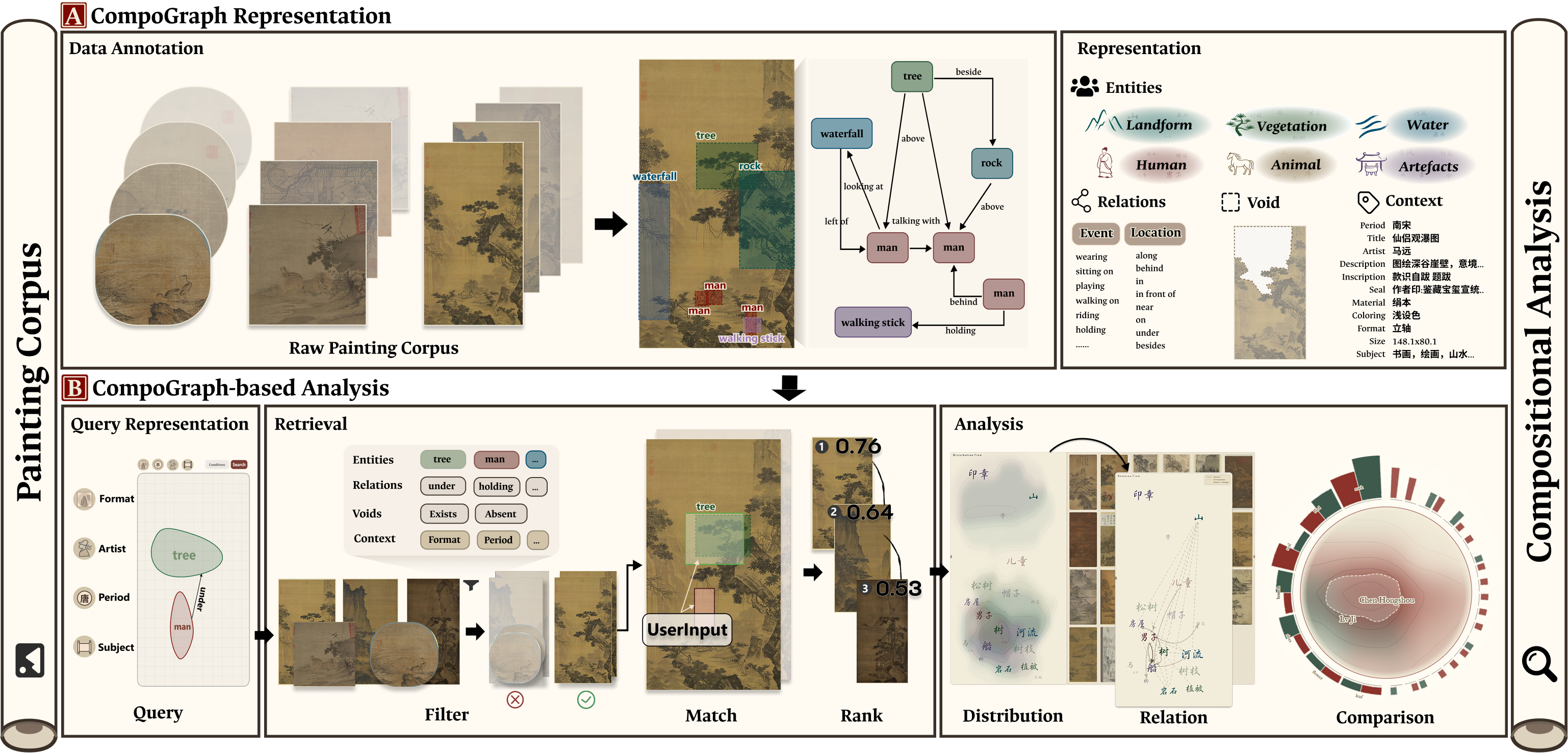}
    \caption{
            Overview of the composition-oriented analysis workflow in \toolname{}.
            Paintings are first represented as \compograph{}s that encode entities,
            relations, voids, and contextual information \paneltag{A}. The
            representation then supports iterative composition querying, candidate
            matching and ranking, cohort construction, and subsequent exploration
            through distribution, relation, and comparison analyses \paneltag{B}.
            }
    \label{fig:pipeline}
\end{figure*}

\textbf{Query representation.}
A query uses the same four-part structure as $G_p$.

\begingroup
\setlength{\abovedisplayskip}{3pt}
\setlength{\belowdisplayskip}{3pt}
\begin{equation}
Q=(E_Q,R_Q,V_Q,C_Q).
\end{equation}
\endgroup


$E_Q$ specifies uniquely identified entities together with their optional spatial regions in a normalized coordinate space defined by the painting format. Each painting is normalized within a canonical coordinate system that preserves the characteristic geometry of its viewing format, since compositional organization depends on the canvas shape. Hanging scrolls are represented in a normalized $1{:}2$ coordinate space, whereas square formats such as album leaves and round fans are represented in a normalized $1{:}1$ space.
$R_Q$ specifies directed source-relation-target relations. $V_Q$ specifies queried voids, and $C_Q$ contains painting-level context filters. The format attribute in $C_Q$ is mandatory because compositional organization and spatial distributions are closely tied to the viewing format, whereas other context attributes can be specified as optional filters. Structured image queries can jointly represent entities, locations, and relations~\cite{johnson2015image,lan2012structured}. We extend this structure with the Void and Context layers of the \compograph{}.

\textbf{Hard filtering.}
Following structured query processing in graph-based retrieval~\cite{pienta2016visage}, the system first applies the compositional conditions specified by the query together with the mandatory format condition and any optional context filters selected by the user. Paintings that lack any queried compositional component or fail to satisfy a specified context filter are excluded. The remaining candidates therefore satisfy the query's hard existence and contextual constraints, while their degree of spatial correspondence is evaluated in the subsequent matching stage.


\textbf{Soft matching and weighted scoring.}
After hard filtering, all remaining candidates satisfy the required entity,
relation, void, and context conditions. Soft matching then evaluates the
spatial correspondence between the query and each candidate. For an entity
constraint $q_j$, the contribution of a candidate entity $x$ is defined as
\begin{equation}
s_{\mathrm{ent}}(q_j,x)=
\max\left\{
\operatorname{IoU}(B_{q_j},B_x),
s_{\mathrm{center}}(q_j,x)
\right\},
\label{eq:entity_contribution}
\end{equation}
where the center-position score is nonzero only when the candidate center
falls within the query region and decreases with its distance from the query
center. For a void constraint $z_j$, which is represented as an irregular
region rather than a bounding box, the contribution is measured by mask IoU:
\begin{equation}
s_{\mathrm{void}}(z_j,z)=
\frac{|Z_{z_j}\cap Z_z|}
{|Z_{z_j}\cup Z_z|}.
\label{eq:void_contribution}
\end{equation}
For each layout constraint, the highest-scoring matching instance or void
region is retained:
\begin{equation}
s_j(Q,p)=
\begin{cases}
\displaystyle\max_{x\in X_j(p)}
s_{\mathrm{ent}}(q_j,x),
& q_j\text{ is an entity constraint},\\[8pt]
\displaystyle\max_{z\in Z(p)}
s_{\mathrm{void}}(z_j,z),
& z_j\text{ is a void constraint}.
\end{cases}
\label{eq:constraint_score}
\end{equation}
The contributions of all layout constraints are then combined using their
user-adjustable weights:
\begin{equation}
S_{\mathrm{layout}}(Q,p)=
\frac{\sum_j\omega_j s_j(Q,p)}
{\sum_j\omega_j},
\label{eq:layout_score}
\end{equation}
where $\omega_j$ denotes the weight of constraint $j$. The default weight is
$1.0$ for each constraint, and users can adjust these weights within the
range $[0,5]$. Candidates are ranked by $S_{\mathrm{layout}}(Q,p)$, and the
top-$K$ candidates are retained as the resulting cohort, where $K$ is
user-adjustable and defaults to $100$. The matched entities and void regions
are preserved as evidence for subsequent inspection.

\begin{figure*}[t]
    \centering
    \includegraphics[width=\linewidth]{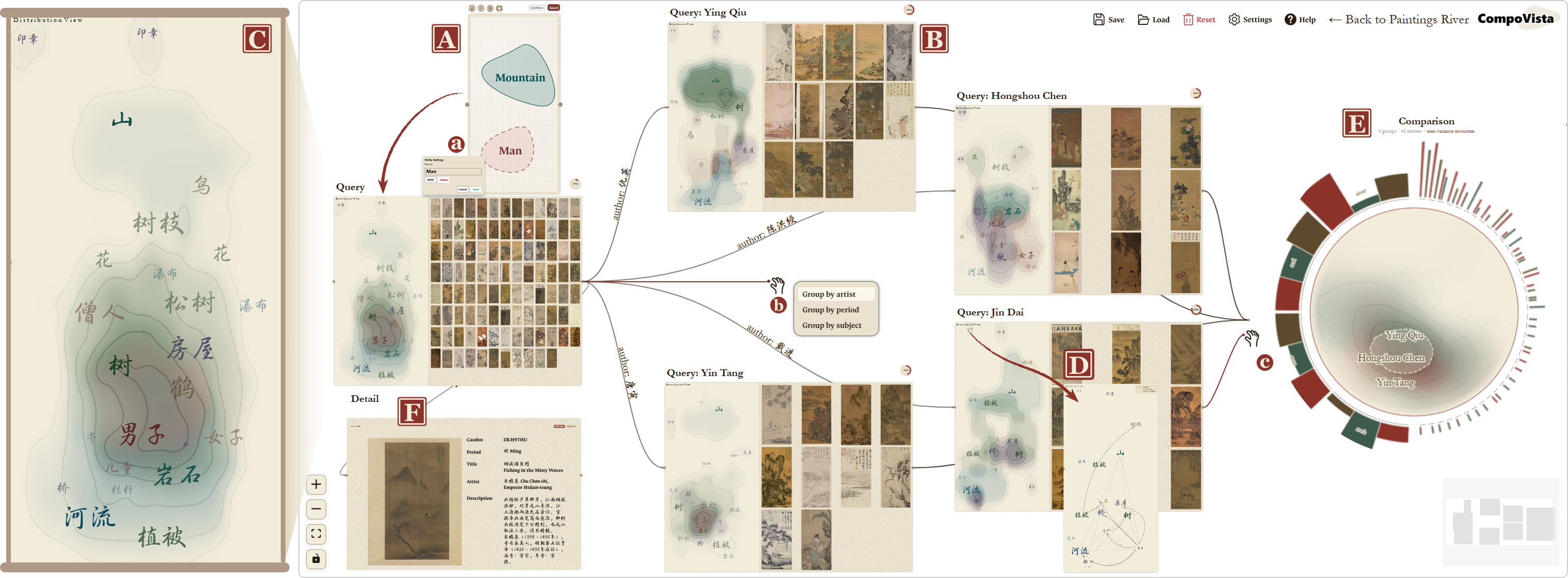}

    \captionof{figure}{
        \toolname{} is a canvas-based tool, enabling art historians to construct and refine composition-oriented cohorts through visual queries, summaries, and evidence inspection.
        Historians use the \syscompInCaption{Query Node}
        \paneltag{A} to express hypotheses by drawing entities
        and connecting relation edges \paneltagRound{a}. Retrieved paintings are
        organized as \syscompInCaption{Cohort Node} \paneltag{B} and explored
        through the \syscompInCaption{Distribution View} \paneltag{C}
        and the \syscompInCaption{Relations View} \paneltag{D}. By dragging
        from a cohort node's right-side port, users can split the cohort
        by artist, period, or subject \paneltagRound{b}; they can also connect a
        cohort node to the \syscompInCaption{Comparison View} \paneltag{E} to add it
        into a comparison \paneltagRound{c}. Finally, users inspect matched
        paintings and supporting evidence in the \syscompInCaption{Detail Node}
        \paneltag{F}.
    }

    \label{fig:teaser}
\end{figure*}


\subsubsection{Cohort Construction and Analysis}

Composition-oriented analysis rarely relies on individual paintings alone. Instead, art historians iteratively organize retrieved paintings into working cohorts, summarize their shared compositional characteristics, compare multiple cohorts, and return to individual paintings to verify their observations. We support this analytical workflow by preserving retrieval evidence with each cohort and enabling cohort construction, summarization, comparison, and painting-level evidence inspection through coordinated visual representations (T2, T3, T5, T6).

\textbf{Cohort construction.}
After identifying relevant paintings, art historians organize retrieved paintings into working cohorts and iteratively refine these cohorts as their analysis progresses. We preserve the retrieval query together with its matching evidence, allowing users to inspect why paintings were retrieved and to revise the query structure or the relative importance of matching criteria. Updated queries trigger the same retrieval and ranking procedure, producing revised cohorts while maintaining the analytical history (T5).

\textbf{Cohort composition summarization.}
A retrieved cohort does not directly reveal recurring compositional patterns or cohort differences. We therefore summarize each cohort using the \compograph{}. Each painting is treated as one observation and every prevalence measure is normalized by the cohort size $|P|$, regardless of repeated instances, following painting-level aggregation~\cite{glueck2017phenostacks,somarakis2021visual}. Entity prevalence is defined as the proportion of paintings containing a given entity category. Spatial prevalence aligns entity bounding boxes within a format-normalized coordinate space, where each grid cell records the proportion of paintings overlapped by at least one instance of the selected category, following aligned bounding-box aggregation~\cite{deng2023visimages}. Relation support is computed as the proportion of paintings containing the same directed source--type--target relation, while co-occurrence support records the proportion containing both entity categories regardless of whether an explicit semantic relation is annotated~\cite{agrawal1993mining}. Void prevalence is computed in the same normalized coordinate space, where each cell records the proportion of paintings containing an overlapping void region. These summaries capture recurring compositional characteristics within the cohort and serve as the basis for subsequent comparison (T2).

\textbf{Cohort comparison and evidence checking.}
For two cohorts, we compare the prevalence of each pattern. A pattern is more common in the cohort with the higher prevalence, and the difference between the two values shows the size of the contrast~\cite{somarakis2021visual}. Entity and relation differences use their cohort proportions. Spatial and void differences are computed cell by cell. For more than two cohorts, we use the range between the highest and lowest prevalence and record which cohorts produce them. All comparisons use the original painting-level prevalence values between 0 and 1. The results are descriptive, not inferential. Each result remains linked to its supporting paintings and annotations~\cite{somarakis2021visual} (T3, T6).

\section{\toolname{}}
\label{sec:system}
We present \toolname{}, a canvas-based interactive tool to assist researchers in implementing the methods mentioned in Sec.~\ref{sec:method}. This section provides an overview of the system and details of visual design and interaction.

\subsection{System Overview}
\label{sec:system_overview}
\toolname{} is built on the dataset described in Sec.~\ref{sec:method}.
The interface is organized as a free-form analytical canvas (Fig.~\ref{fig:teaser}). Art historians start from the \syscomp{Query Node} (Fig.~\ref{fig:teaser} \paneltag{A}) to express composition-oriented questions through visual queries and context filters. The backend matches the query against the corpus and returns results as a \syscomp{Cohort Node} (Fig.~\ref{fig:teaser} \paneltag{B}). Users can then examine the cohort through the \syscomp{Distribution View} (Fig.~\ref{fig:teaser} \paneltag{C}) and the \syscomp{Relations View} (Fig.~\ref{fig:teaser} \paneltag{D}), compare multiple cohorts in the \syscomp{Comparison View} (Fig.~\ref{fig:teaser} \paneltag{E}), and inspect painting-level evidence in the \syscomp{Detail Node} (Fig.~\ref{fig:teaser} \paneltag{F}). By linking query, retrieval, cohort analysis, comparison, and evidence inspection on the same canvas, \toolname{} supports an iterative workflow from hypothesis to painting-level verification.

\begin{figure*}[t]
    \centering
    \includegraphics[width=\textwidth]{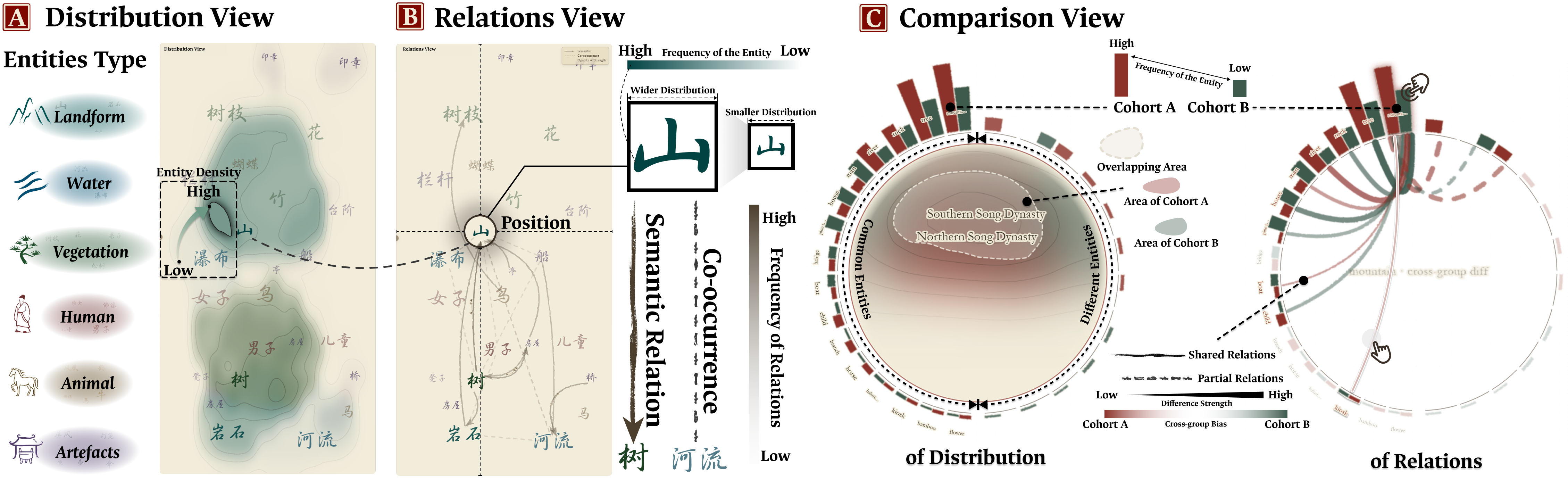}
    \caption{
    Visual designs for cohort-based analysis.  The views encode cohort distributions \paneltag{A}, within-cohort spatial and relational patterns \paneltag{B}, and
    cross-cohort spatial and relational differences \paneltag{C}.
    }
    \label{fig:visual_design}
\end{figure*}

\subsection{Visual Design}
\label{sec:visual_design}

To support cohort-level compositional analysis, \toolname{} provides
three coordinated visual summaries: the \syscomp{Distribution
View} (Fig.~\ref{fig:visual_design} \paneltag{A}), the
\syscomp{Relations View} (Fig.~\ref{fig:visual_design}
\paneltag{B}), and the \syscomp{Comparison View}
(Fig.~\ref{fig:visual_design} \paneltag{C}). These views summarize a cohort from three
complementary perspectives: where entities tend to appear, which
entities tend to appear together, and how compositional patterns differ
across cohorts.

\begin{figure}[t]
    \centering
    \includegraphics[width=\columnwidth]{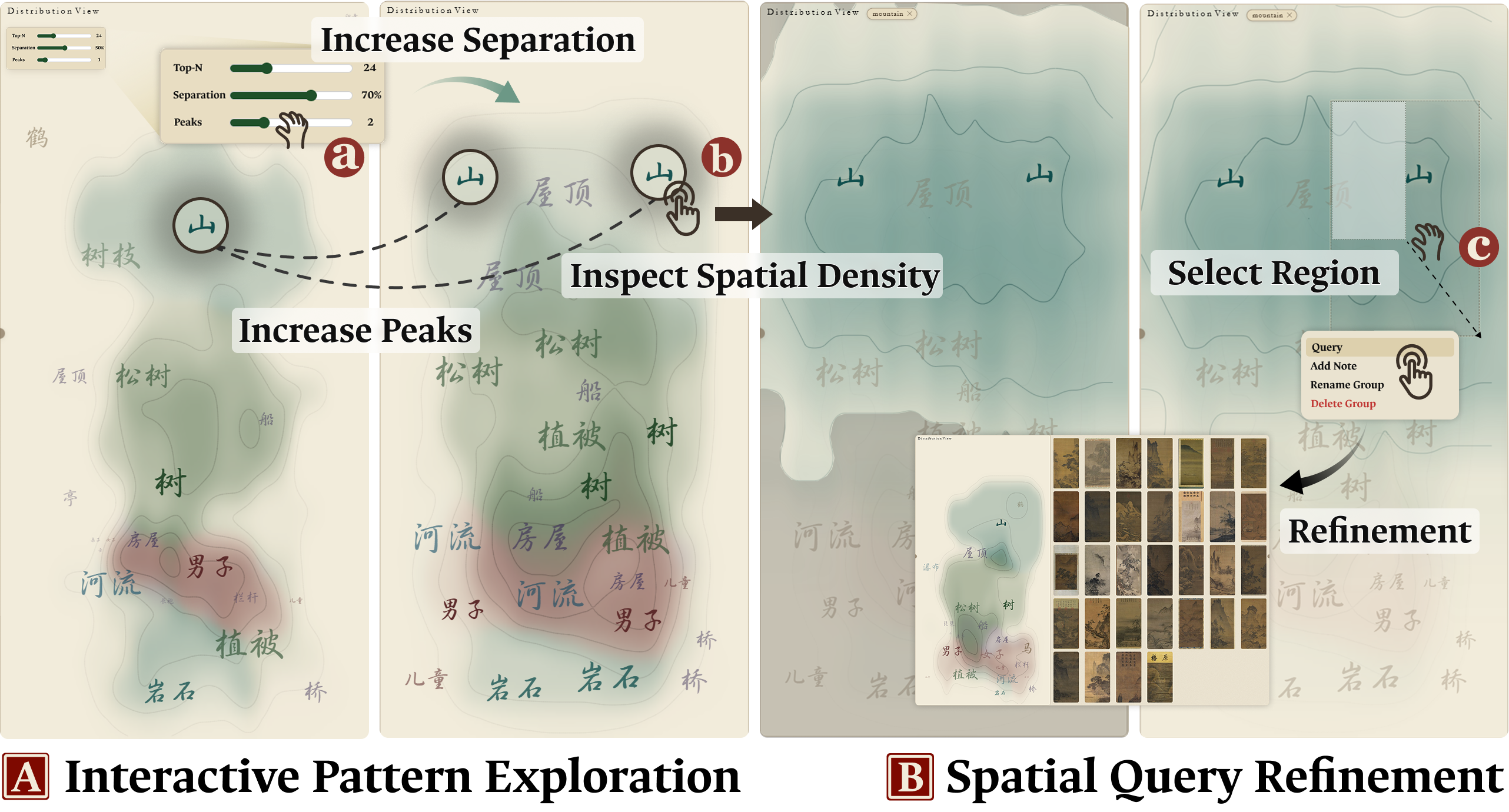}
    \caption{
    Interactive exploration supported by the \syscompInCaption{Distribution View}.
    Users refine visual patterns through parameter adjustment
    \paneltag{A}, and perform region-based spatial queries for iterative cohort refinement \paneltag{B}.
    }
    \label{fig:interaction}
\end{figure}

\subsubsection{Distribution View}

Motivated by \textbf{T2}, the \syscomp{Distribution View} supports overview inspection of entity distributions while preserving spatial context. Following the overview-first principle of visual information seeking~\cite{shneiderman1996eyes}, the view provides an aggregated summary of a painting cohort that allows users to identify dominant compositional patterns before inspecting individual paintings. The view places entities in a format-normalized coordinate space, where each word represents an entity. Position is used as the primary visual channel because it provides accurate perceptual encoding for spatial comparison~\cite{cleveland1984graphical}. Each entity is located at its average position within the cohort, while the surrounding region indicates its spatial spread. To support iterative exploration (Fig.~\ref{fig:interaction}), users can interactively adjust layout constraints through a control panel to explore different distribution patterns \paneltagRound{a}, inspect the full spatial density of individual entities \paneltagRound{b}, and perform region-based spatial queries for cohort refinement \paneltagRound{c}. Color encodes entity category, including \entityHumanTagText{human
figures}, \entityVegetationTagText{vegetation},
\entityLandformTagText{landform}, \entityWaterTagText{water},
\entityAnimalTagText{animal}, and \entityArtefactTagText{artefact}. The
palette follows perceptual color design principles~\cite{zeileis2009escaping}
and also reflects domain intuition in \TCP{}s. 
For example, the landform
color follows the blue-green mountain tone associated with
\textit{A Thousand Li of Rivers and Mountains}, while water and vegetation
use familiar blue and green hues.
Unlike conventional word clouds, where font size usually represents term
frequency~\cite{felix2017taking}, the \syscomp{Distribution View} uses
font size to encode the average spatial size of an entity and opacity to
encode its occurrence frequency. 
This separates two things that are often
mixed together in compositional analysis: how much pictorial space an
entity occupies, and how often it appears across the cohort. 
For example,
a small human figure may appear often but may not dominate the pictorial
structure. Mountains, rocks, trees, or water may define the composition
because they occupy larger parts of the painted surface. To fit the \TCP
context, entity words use a calligraphic style inspired by
\textit{Wang Xizhi}'s handwriting, and the background uses a paper-like
texture reminiscent of Xuan paper, a traditional Chinese paper widely used in ink and brush painting. These choices help the aggregated
layout read like a pictorial composition while keeping the spatial
evidence inspectable.

\subsubsection{Relations View}

Motivated by \textbf{T2}, the
\syscomp{Relations View} supports the inspection of recurring
compositional relations while preserving the spatial context established
in the \syscomp{Distribution View}. To facilitate comparison across
views, the design follows the principle of mental map
preservation~\cite{misue1995layout}, allowing users to shift attention from entity distributions to relations without reconstructing the spatial layout.
The view reuses the same format-normalized coordinate space as the \syscomp{Distribution View}. Entities retain their positions, font sizes, opacities, and category colors, while edges are added to
represent relations between them. This coordinated design enables
users to inspect relations while maintaining the compositional context
established in the distribution overview~\cite{baldonado2000guidelines}.
Edges represent associations between
entities. Directed solid edges represent annotated semantic relations,
such as \textit{sitting on} or \textit{holding}. Dashed edges represent
co-occurrence relations, showing how often two entities appear together.
Edge opacity encodes relation support, with darker lines showing more
frequent semantic relations or co-occurrences within the cohort. By
preserving the spatial layout and highlighting relations, the
\syscomp{Relations View} helps users identify stable compositional
configurations. It is also useful for motif-related analysis, because
such questions often depend on relations among entities, not only on the
entities themselves.

\subsubsection{Comparison View}

Motivated by \textbf{T3}, the
\syscomp{Comparison View} supports cross-cohort comparison by combining
entity statistics, spatial distributions, and relation patterns in a
coordinated view, enabling users to examine both global trends and
detailed compositional differences~\cite{shneiderman1996eyes}. The view adopts a radial ring-based layout in which entities occupy
fixed angular positions across cohorts, providing a shared reference
frame for comparison~\cite{krzywinski2009circos}. Colored radial bars
encode normalized entity prevalence. The ring is partitioned into shared
and cohort-specific regions: entities present in all cohorts are placed
in the left semicircle, whereas those absent from one or more cohorts
are placed in the right semicircle, making common patterns and
structural differences easy to distinguish. When an entity is selected, the center displays a spatial comparison map computed from normalized density fields across cohorts. Cohort-specific regions retain their colors, whereas overlapping regions are emphasized through color blending, translucent masks, and dashed contours to reveal shared spatial patterns. Users can also inspect individual cohort distributions by hovering over cohort labels. Relation differences are visualized as curved chords~\cite{krzywinski2009circos}. Chord thickness encodes the difference in relation strength, opacity indicates relation support, and color denotes the cohort in which the relation is most dominant. Solid chords represent shared relations, whereas dashed chords indicate cohort-specific associations. Together, these coordinated encodings support comparison across entity prevalence, spatial placement, and relational structure~\cite{gleicher2011visual}.

\subsection{Interactions}
\label{sec:interaction_design}

\toolname{} supports interactions that help art historians construct,
branch, compare, and inspect composition-oriented cohorts on the analytical
canvas~\cite{heer2012interactive,shneiderman1996eyes}. Basic operations such as panning, zooming, node dragging, and
node opening are supported throughout the workspace.

\textbf{Construct visual queries.}
Art historians can formulate a compositional hypothesis directly on the
\syscomp{Query Node}~\cite{ahlberg1992dynamic,shneiderman2002dynamic} (Fig.~\ref{fig:teaser} \paneltagRound{a}).
They add entity regions from the controlled vocabulary, drag or resize them to specify approximate locations and spatial extents, mark voids when needed, and connect entities with relation edges. 
Contextual filters such as artist, period, subject, or format can be combined with the visual query. These contextual conditions remain attached to the resulting cohort, helping users interpret later summaries and comparisons in relation to the art-historical context in which the cohort was created.

\textbf{Branch cohorts by context.}
To examine how a cohort varies across historical or contextual
conditions, users drag from the right-side port of a
\syscomp{Cohort Node} (Fig.~\ref{fig:teaser} \paneltagRound{b}). The system
then lets them split the cohort by artist, period, or subject, producing
child cohort nodes linked to the original group. This interaction keeps
the analysis history visible on the canvas and allows users to move
from a broad retrieved group to more specific subgroups without
restarting the search.

\textbf{Add cohorts to comparison.}
Users can also drag from the right-side port of a \syscomp{Cohort Node}
to a \syscomp{Comparison View} (Fig.~\ref{fig:teaser} \paneltagRound{c}).
The selected cohort is then added to the comparison context. Additional
cohorts can be connected in the same way, enabling art historians to build
pairwise or multi-group comparisons incrementally while preserving the
links between compared groups and their source queries.

\textbf{Inspect painting-level evidence.}
Users can select one or more paintings within a \syscomp{Cohort Node}, invoke a context menu, and open the \syscomp{Detail Node} (Fig.~\ref{fig:teaser} \paneltag{F}) to inspect the selected paintings and supporting evidence. The
node shows the painting, matched entities, supporting relations,
voids, and contextual information. This helps users check
whether a cohort-level pattern is supported by the actual paintings.

\subsection{Implementation}
\label{sec:implementation}

\toolname{} is built with JavaScript (Vue 3 and D3.js) on the frontend and Python (FastAPI) on the backend. The frontend uses Vue Flow~\cite{vueflow2025} to support the
node-based analytical workspace, with state management and interface
components built using standard Vue ecosystem libraries, and
D3.js~\cite{bostock2011d3} for custom visualizations. The backend
stores painting context, \compograph{} annotations, void representations, and query provenance, and serves the matching and
aggregation results used by the coordinated views.

\section{Evaluation}
\label{sec:eval}

We evaluated \toolname{} through two representative case studies and
a user study with 12 art historians, followed by semi-structured interviews.
Together, these evaluations assess the usefulness and usability of \toolname{} for composition-oriented analysis of \TCP{}s.



\subsection{Case Studies}

We present two representative case studies 
reconstructed from the user study described in Sec.~\ref{sec:user_study}. Drawing on participants' interaction logs, screen recordings, and task notes, the cases
illustrate two different analytical workflows: (1) constructing and comparing painting cohorts to identify recurring compositional patterns, and (2) starting from compositional structure to iteratively retrieve and refine painting cohorts.

\begin{figure*}[t]
    \centering
    \includegraphics[width=\textwidth]{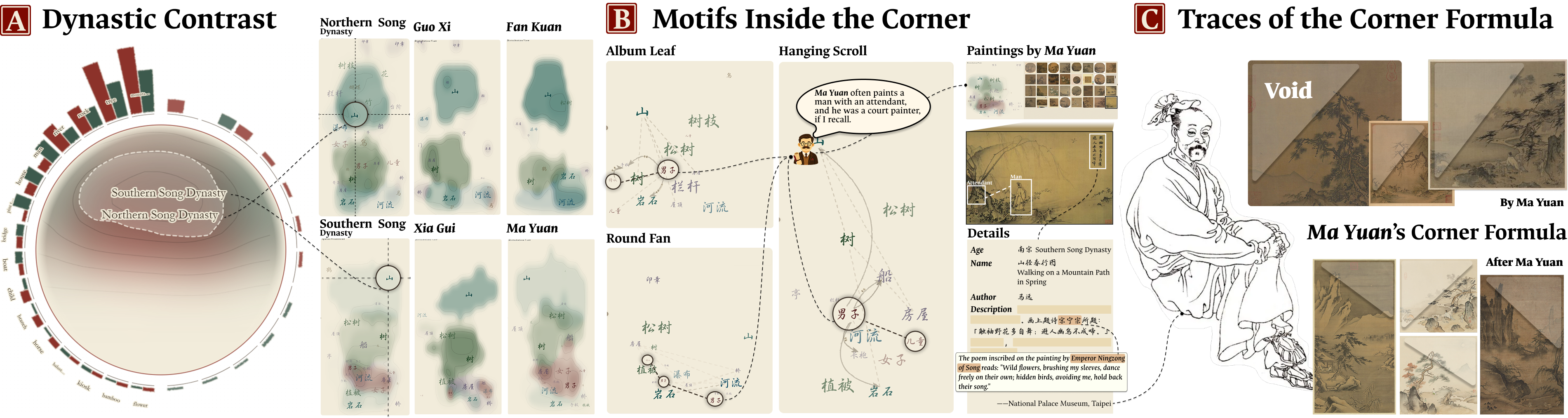}
    \caption{
    Case 1 exploration process. 
    \protect\paneltag{A} \textbf{P3} first compares Northern and Southern Song layouts and finds a shift from the center to the corner. 
    \protect\paneltag{B} \textbf{P3} then explores \textit{Ma Yuan}'s cohort, where a recurring man-attendant pattern suggests a courtly context. 
    \protect\paneltag{C} Through structural querying, \textbf{P3} finds that the corner formula was reused in later paintings.
    }
    \label{fig:case_study1}
\end{figure*}

\paragraph{Case 1. From Center to Corner}
\label{sec:case_song_maxia}

\textbf{P3} is an art historian specializing in Song Dynasty landscape painting, 
with a particular interest in the compositional differences between 
Northern and Southern Song landscapes.
Using \toolname{}, \textbf{P3} compared painting cohorts from the two periods and traced how distinctive compositional patterns emerged within them.

\textbf{Cohort construction and distribution overview.}
\textbf{P3} first created two cohorts of hanging-scroll paintings from the
Northern and Southern Song periods, respectively, and arranged them vertically for comparison. In the \syscomp{Distribution
View}, the Northern Song cohort showed a centered and relatively
symmetric distribution of \entityLandformTagText{mountain},
\entityVegetationTagText{tree}, and \entityLandformTagText{rock}, whereas
the Southern Song cohort showed greater concentrations of these entities toward the lower and lateral regions, particularly the corners
(Fig.~\ref{fig:case_study1}). \textbf{P3} commented, ``\textit{Given the different political and cultural
conditions of the Northern and Southern Song, such a compositional
difference is not unexpected. However, this was the first time I could
see the contrast so directly across a large collection of paintings}''
(\textbf{T2, T5}).

\textbf{Cohort comparison and artist-based branching.}
To examine the dynastic difference more closely, \textbf{P3} connected the
Northern Song and Southern Song cohorts to the
\syscomp{Comparison View} (Fig.~\ref{fig:case_study1} \paneltag{A}) (\textbf{T3}). After selecting
\entityLandformTagText{mountain}, the spatial comparison showed that
both cohorts included mountains in the central region,
but their dominant placements differed. Mountains were concentrated around the
pictorial center in Northern Song paintings, but appeared more frequently in lateral and corner regions in Southern Song paintings. The contrast therefore reflected a redistribution of compositional weight rather than a simple presence-or-absence difference. \textbf{P3} then branched the two cohorts by artist to examine this redistribution in greater detail (\textbf{T5}). In the Northern Song cohort, paintings by \textit{Guo Xi} and \textit{Fan Kuan} showed a strong
tendency toward central mountain organization. In the Southern Song cohort, \textit{Ma Yuan} and \textit{Xia Gui} both exhibited less centralized arrangements
, but in different ways: \textit{Ma Yuan} concentrated
mountains, rocks, trees, and small \entityHumanTagText{human figures}
in one corner, whereas \textit{Xia Gui} extended landscape elements along one
side. \textbf{P3} noted that \textit{Ma Yuan} pushed the edge-oriented tendency toward a corner formula, while \textit{Xia Gui} developed it as a lateral, side-stretched organization~\cite{fong1973sung,barnhart1997three}.

\textbf{Relation discovery and context exploration.}
Since \textit{Ma Yuan}'s pattern was especially pronounced, \textbf{P3} retrieved
all paintings attributed to him in the corpus and opened them as a new 
cohort (\textbf{T4}). \textbf{P3} then inspected the \syscomp{Relations
View} (Fig.~\ref{fig:case_study1} \paneltag{B}) to understand what entities repeatedly appeared within \textit{Ma Yuan}'s works. Besides expected landscape elements such as \entityLandformTagText{mountain}, \entityLandformTagText{rock}, 
\entityVegetationTagText{tree}, and \entityWaterTagText{river}, 
\entityHumanTagText{man} appeared as a frequent entity, often together 
with an \entityHumanTagText{attendant} or other small figure elements
(\textbf{T2}). This recurring \entityHumanTagText{man}--\entityHumanTagText{attendant} association shifted \textbf{P3}'s attention from 
spatial composition alone to the human scenes embedded within \textit{Ma Yuan}'s landscapes. \textbf{P3} commented, ``\textit{I remember that Ma Yuan produced paintings for the Southern Song court, so these small figures may not be merely incidental staffage.}''  Motivated by this observation, \textbf{P3} opened
the \syscomp{Detail Node} and inspected related contextual materials (\textbf{T1, T6}). There, \textbf{P3} found references to imperial inscriptions and courtly album paintings, including \textit{Songdi Mingti Ce}. This reminded \textbf{P3} of a common interpretation of the album: the small figure may represent a Song emperor rather than an ordinary staffage figure~\cite{murck1991words,fong1973sung}. \textbf{P3} therefore treated the recurring \entityHumanTagText{man}--\entityHumanTagText{attendant} pattern as an
interpretive clue that \textit{Ma Yuan}'s corner formula could carry a courtly
viewing context, while noting that this remained a hypothesis requiring
further art-historical evidence.

\textbf{Structural query and formula retrieval.}
Finally, \textbf{P3} abstracted \textit{Ma Yuan}'s corner formula
into a structural query (\textbf{T4}) comprising dense \entityLandformTagText{mountain}, \entityLandformTagText{rock}, and \entityVegetationTagText{tree} elements in one corner, and a large void in the opposite region. \textbf{P3} cloned the query across
the other three corners, and searched the full corpus without restricting
the author to \textit{Ma Yuan}. Most high-ranked results were works
by \textit{Ma Yuan}, but several works by other artists shared the same compressed corner structure and were described in context as following, imitating,
or being in the manner of \textit{Ma Yuan} (\textbf{T1, T6}). \textbf{P3} therefore read them not as false positives, but as traces of how \textit{Ma Yuan}'s formula was reused. ``\textit{The system let me move from period style to author style, and finally to a reusable formula,}'' \textbf{P3} concluded. ``\textit{The interesting part is that the formula was not only spatial; in Ma Yuan's works, it could also carry a small courtly scene.}''

Throughout the analysis, \textbf{P3} iteratively constructed five major
cohorts, involving 383 paintings, 2,817 annotated entities, and 1,786
annotated relations. These cohorts were progressively refined through
comparison, branching, and structural retrieval.

\paragraph{Case 2. Fisherman Postures and River Composition}
\label{sec:case_fisherman}

\textbf{P4} used \toolname{} to examine how this theme appears across formats and what visual structures recur within it.

\begin{figure}[t]
    \centering
    \includegraphics[width=\columnwidth]{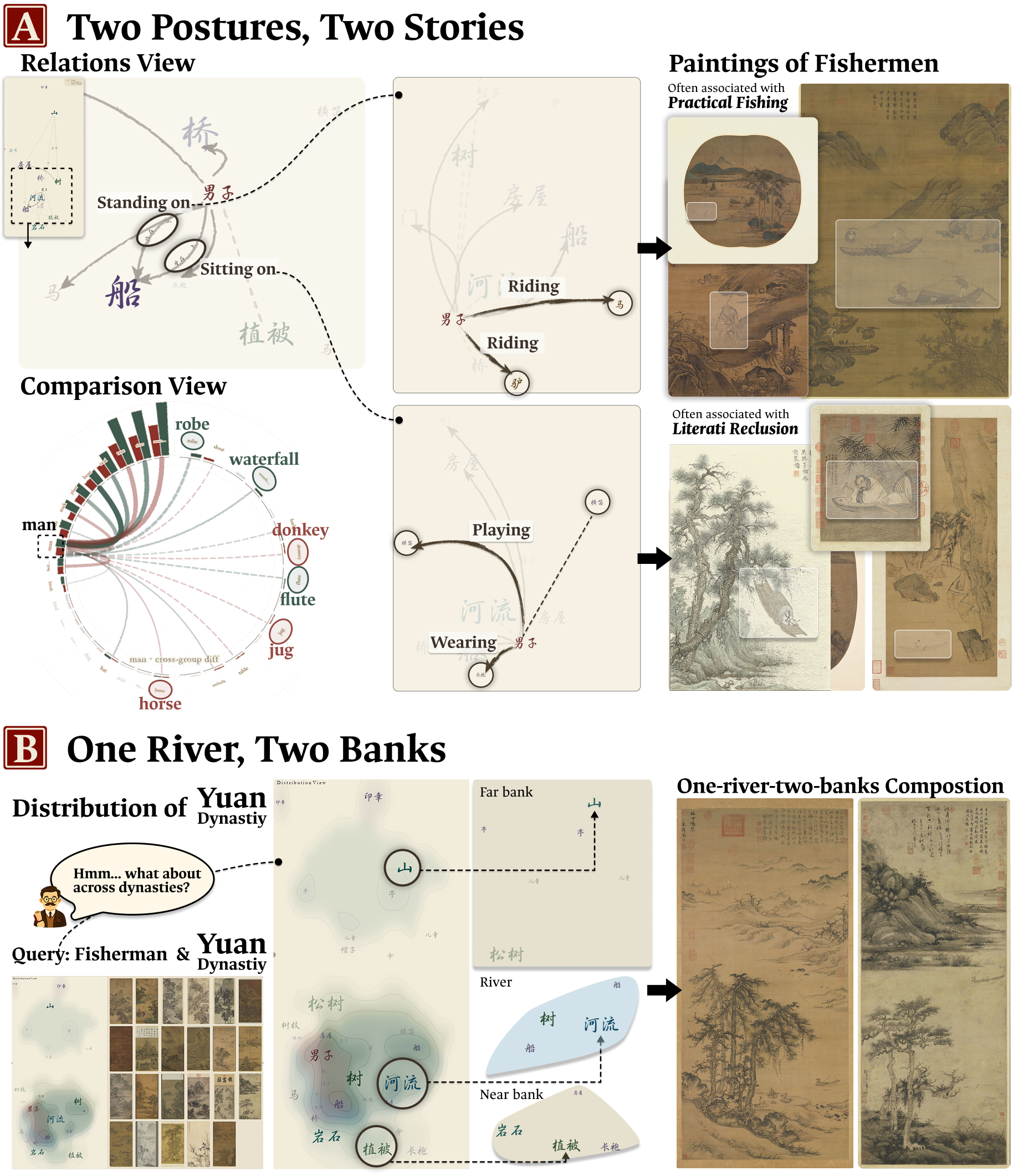}
    \caption{
    Case 2 exploration process. 
    \protect\paneltag{A} \textit{Two Postures, Two Stories} shows how 
    relation-level comparison separates practical fishing scenes from 
    literati reclusion scenes.
    \protect\paneltag{B} \textit{One River, Two Banks} shows how 
    dynastic distribution analysis reveals a Yuan river-centered 
    composition organized by water and opposing banks.
    }
    \label{fig:case_study2}
\end{figure}

\textbf{Fisherman cohort construction.}
\textbf{P4} first searched the keyword \textit{fisherman}. The system grouped
matched paintings by format, including hanging scrolls, album leaves,
and round fans (\textbf{T2, T4}). Because hanging scrolls formed the largest
group, \textbf{P4} opened this cohort. The \syscomp{Distribution View} did not
immediately draw \textbf{P4}'s interest, so \textbf{P4} turned to the
\syscomp{Relations View} to examine the theme's entity-relation
structure (Fig.~\ref{fig:case_study2} \paneltag{A}).
The view revealed a stable core around the fisherman motif:
\entityHumanTagText{man}, \entityArtefactTagText{boat} and \entityWaterTagText{river} appeared
frequently, together with two salient relations, \textit{sitting on
boat} and \textit{standing on boat}. \textbf{P4} commented, ``\textit{The fisherman is
not just a person in the painting; the theme comes from the relation
among the person, the boat, and the water}'' (\textbf{T2}).

\textbf{Comparing sitting and standing fishermen.}
\textbf{P4} returned to the \syscomp{Query Node} and constructed two
relation queries, \textit{sitting on boat} and \textit{standing on
boat}, which produced two comparable cohorts (\textbf{T4}). \textbf{P4}
connected the two cohorts to the \syscomp{Comparison View} to compare
their entity frequencies and relation patterns (\textbf{T3}). The difference
view showed that the two cohorts differed not only in entity prevalence,
but also in the strength of relations among entities. This suggested that
\textit{sitting on boat} and \textit{standing on boat} were associated
with different visual contexts. To understand what these relation differences meant, \textbf{P4} opened the
\syscomp{Relations View} for inspection.
In the \textit{sitting on boat} cohort, the figure
was often associated with playing a flute, wearing a long robe, or
being accompanied by an attendant. After opening the
\syscomp{Detail Node} and inspecting the matched paintings, \textbf{P4} found that these relations corresponded to richer boat-based life scenes.
By contrast, the \textit{standing on boat} cohort showed fewer such
secondary relations and was more closely tied to movement around the boat
and water. \textbf{P4} commented, ``\textit{This is quite interesting. Although both groups depict fishermen, a subtle difference in the figure's posture seems to
change the meaning of the scene. The seated figure conveys a more
contemplative and literati-oriented mood, whereas the standing figure
appears more connected to everyday activity along the river}''
(\textbf{T2, T6}).

\textbf{Dynastic layout contextualization.}
After examining these semantic differences, \textbf{P4} returned to the full
fisherman cohort and branched it by period (\textbf{T5}). 
In the \syscomp{Distribution View}, the Yuan Dynasty group showed a distinctive layout: \entityWaterTagText{river}
occupied the central pictorial space, while \entityLandformTagText{rock},
\entityVegetationTagText{plant}, and other shore elements appeared below
or along the banks (Fig.~\ref{fig:case_study2} \paneltag{B}). \textbf{P4} connected this with
the ``\textit{one river, two banks}'' compositional scheme~\cite{cahill1976hills,fong1973sung}, where the river organizes the
painting by separating banks and positioning human activity within a
broader water-and-shore structure. This helped \textbf{P4} reinterpret the Yuan
Dynasty group as a river-centered composition, where the fisherman motif
was organized by water, banks, and human activity rather than by the
figure alone (\textbf{T2}).

Throughout the analysis, \textbf{P4} iteratively constructed three major cohorts, involving 112 paintings, 947 annotated entities, and 703 annotated relations.

\subsection{User Study}
\label{sec:user_study}

To evaluate whether \toolname{} supports composition-oriented analysis in a usable workflow, we conducted a user study with domain participants. The study focused on whether participants with relevant backgrounds could understand the system workflow, complete composition-oriented tasks, and perceive the system as useful for exploring compositional structures in \TCP{}s. 

\begin{figure*}[t]
    \centering
    \includegraphics[width=\textwidth]{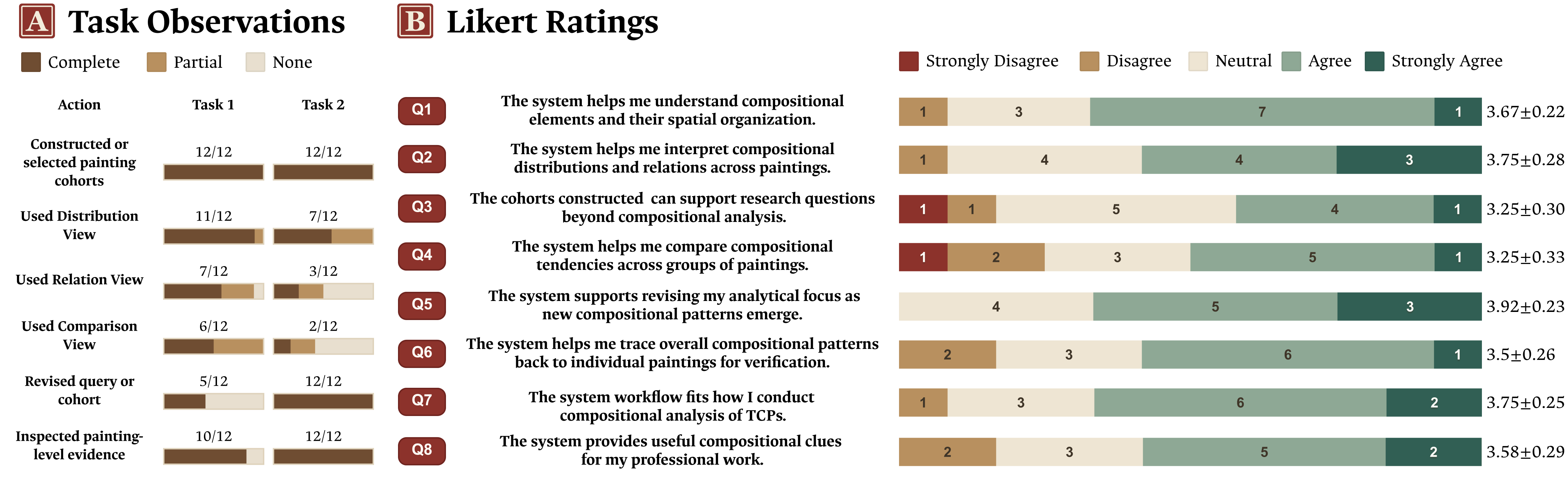}
    \caption{
    Results of the user study. 
    \paneltag{A} Observed analytical actions performed by participants during the two tasks. 
    \paneltag{B} Likert-scale ratings on structure understanding, cohort analysis, iterative evidence checking, and perceived workflow fit and professional value.
    }
    \label{fig:UserStudy}
\end{figure*}

\textbf{Participants.}
We recruited 12 participants with backgrounds in Chinese painting, art history, cultural heritage analysis or related fields. These participants were separate from the two art historians who contributed to the formative study. Among them, 7 were female and 5 were male. 
All participants possessed substantial expertise in \TCP through academic research, professional teaching, or public dissemination. Among them, P1–P2 are university professors specializing in Art History, P3–P4 are associate professors in Art History, and P5–P9 are lecturers in Chinese Painting. 
P10--P12 are museum professionals engaged in research, documentation, and public interpretation of \TCP{}s.
Each participant received compensation in RMB equivalent to USD 20 per hour.

\textbf{Procedure.}
Each user study session was conducted individually, either in person or remotely through online screen sharing. 
In total, 7 sessions were conducted in person and 5 sessions were conducted via screen sharing. Each session lasted approximately 90 minutes. 
The session began with a 10-minute introduction to the study background and dataset (Sec.~\ref{sec:dataset}), followed by a 5-minute tutorial introducing the main components and interactions of \toolname{}, during which participants familiarized themselves with the interface through several example queries.
Participants then completed two analysis tasks, each taking approximately 15 minutes.
The purpose of Task~1 was to examine whether \toolname{} could help users understand and compare compositional structures across painting cohorts. In this task, participants constructed or selected comparable painting cohorts and compared their compositional patterns. The purpose of Task~2 was to investigate whether \toolname{} could support exploratory compositional analysis. Participants iteratively retrieved and refined painting cohorts based on user-defined visual queries. During both tasks, we encouraged participants to think aloud and recorded their interactions, task notes, and verbal feedback.
We further coded participants' engagement in key analytical behaviors based on screen recordings, interaction logs, and session observations. Following prior work that abstracts low-level user activities into higher-level analytical behaviors in exploratory visual analysis~\cite{blascheck2016visual,reda2016modeling,guo2015case}, we developed a three-level ordinal coding scheme tailored to our analytical workflow. \textit{Complete} indicated that a participant carried out the behavior and used the corresponding view or evidence to support analysis; \textit{Partial} indicated that the participant initiated or attempted the behavior but used it incompletely or superficially; and \textit{None} indicated that the behavior was not observed. The session concluded with a 45-minute post-study phase, during which participants completed an eight-item 5-point Likert-scale questionnaire, followed by a semi-structured interview. The questionnaire covered four aspects: structure understanding (Q1--Q2), cohort analysis (Q3--Q4), iterative evidence checking (Q5--Q6), and perceived workflow fit and professional value (Q7--Q8).



\textbf{Results.}
Fig.~\ref{fig:UserStudy} summarizes both task observations and questionnaire results. As shown in Fig.~\ref{fig:UserStudy} \paneltag{A}, participants generally used the system according to the intended compositional analysis workflow: all participants constructed or selected painting cohorts in both tasks (12/12), most inspected painting-level evidence (10/12 in Task~1 and 12/12 in Task~2), and many used the \syscomp{Distribution View} (11/12 in Task~1 and 7/12 in Task~2). Query or cohort refinement was especially common in Task~2 (12/12), suggesting that participants used the system iteratively as they refined questions and checked evidence. Fig.~\ref{fig:UserStudy} \paneltag{B} shows a similar pattern in the questionnaire results. Participants generally gave positive ratings to \toolname{}. Most participants agreed or strongly agreed that the system helped them understand compositional elements and their spatial organization (Q1, 8/12), supported revising their analytical focus as new compositional patterns emerged (Q5, 8/12), and fit their compositional analysis workflow (Q7, 8/12). More than half also agreed or strongly agreed that it helped them interpret compositional distributions and relations across paintings (Q2, 7/12), trace overall compositional patterns back to individual paintings for verification (Q6, 7/12), and provide useful compositional clues for professional work (Q8, 7/12). Responses were more mixed for organizing painting cohorts around broader research questions (Q3, 5/12 agreed or strongly agreed) and comparing compositional tendencies across painting cohorts (Q4, 6/12 agreed or strongly agreed). Together, these results suggest that \toolname{} was most effective in supporting structured composition understanding and iterative analytical exploration, while broader research-oriented uses of the constructed cohorts received more mixed evaluations.

\subsection{Expert Interviews}
\label{sec:expert_interview}


We next summarize the qualitative findings from the post-study interviews. We analyzed the interview transcripts and notes using thematic analysis~\cite{braun2006using}. The interviews focused on three aspects: whether the system helped participants construct and revise composition-oriented cohorts, whether the spatial distributions, relational structures, and painting-level evidence supported compositional interpretation, and what limitations or improvements they identified for research or teaching use. The findings are organized into three themes: perceived analytical value, workflow usability, and limitations and suggestions.


\textbf{Perceived analytical value.}
Participants generally agreed that \toolname{} helped transform broad art-historical interests into inspectable compositional evidence. \textbf{P5} noted that the cohort-based workflow made it easier to ``\textit{first see what kinds of paintings are being compared.}'' \textbf{P8} explained that the \syscomp{Distribution View} helped distinguish recurring patterns from isolated examples: ``\textit{The distribution view gives me a first sense of the pattern.}'' At the same time, \textbf{P10} emphasized the importance of validating overview patterns through individual paintings: ``\textit{The overview is useful, but I still need to click back to the actual paintings. Only after seeing the examples can I judge whether the pattern really makes sense.}'' Together with the workflows reported in the case studies, these comments indicate that \toolname{} supports hypothesis development by linking cohort construction, overview analysis, comparison, and painting-level inspection.

\textbf{Workflow usability.}
Participants found the workflow consistent with art-historical analytical practice. \textbf{P2} noted that keeping each cohort as a node helped them remember ``\textit{which group I am looking at and why it was created.}'' \textbf{P6} valued the link between overview summaries and painting-level evidence: ``\textit{I do not want to stop at a statistical pattern. I want to see the paintings behind it and check whether the visual evidence is convincing.}'' Participants showed different entry preferences. \textbf{P9} preferred the \syscomp{Distribution View} when examining landscape compositions. In contrast, \textbf{P11} found the \syscomp{Relations View} more natural for motif-based questions because it highlighted recurring entity associations: ``\textit{For motif questions, the relation between things is often more important than the entity itself. Seeing which entities repeatedly appear together helps me think about the scene.}'' Consistent with the case studies, \textbf{P3} benefited from spatial summaries for landscape comparison, whereas \textbf{P4} used relational evidence to understand the fisherman motif. \uline{These observations suggest that \toolname{} should support multiple paths into compositional analysis rather than assuming a single fixed workflow.}

\textbf{Limitations and suggestions.}
Experts also identified several directions for improvement. \uline{First, the system should explain more clearly why particular paintings are retrieved or ranked, especially when a match depends on partial spatial similarity.} As \textbf{P7} noted, ``\textit{Although I can see that the result is related to the query, I still want to know which part of the painting makes it match the query. This would help me conduct a deeper analysis of the relevant part.}'' \uline{Second, the system should support more abstract and fuzzy composition queries.} Experts pointed out that art historians often reason about approximate regions, visual balance, density, openness, or compositional tendencies rather than exact entity locations. \textbf{P1} explained that experts often think in terms of ``\textit{a general feeling of balance, density, or openness}'' rather than specific entity positions. Third, the system needs stronger history management to support longer exploration sessions. \textbf{P12} suggested that the system should help scholars ``\textit{return to the groups created at the beginning of the exploration and remember why a comparison was made in the first place.}'' The motivation behind each exploratory step is especially important for sustained research processes.

Experts also noted several limitations. They mentioned a learning cost for interpreting comparison encodings, emphasized that results depend on annotation quality, and suggested caution when comparing paintings across formats. They also noted that handscrolls are currently excluded from normalized spatial analysis because their sequential viewing structure differs from hanging scrolls, album leaves, and round fans.

\section{Discussion}

Based on the evaluation above, this section discusses what \toolname{} suggests for composition-oriented visual analytics of \TCP{}s. We first summarize the key findings, then derive design implications and outline limitations and future work.

\subsection{Key Findings}

\textbf{Inspectable compositional evidence.}
The evaluation suggests that the value of \toolname{} is not to automatically determine art-historical meaning, but to help scholars inspect compositional evidence. Traditional compositional analysis concerns where entities appear, how they relate to each other, how void is organized, and how these structures work within different painting formats~\cite{jiang2012composition,silbergeld1982chinese,fong1992beyond}. Through the \compograph{}, \toolname{} represents entities, relations, void, and context as structured evidence that can be queried, summarized, and inspected.
The two case studies illustrate this role. \textbf{P3} used the system to examine how mountains, rocks, trees, and void were distributed across dynasties and artists, and further connected \textit{Ma Yuan}'s recurring man-attendant relation to imperial inscriptions and courtly album paintings. \textbf{P4} used relations among figures, boats, and water to understand the fisherman motif as a compositional structure, rather than a simple entity label. These cases suggest that \compograph{} representation can support expert interpretation when it remains connected to painting-level visual and contextual evidence.

\textbf{Cohorts as provisional evidence spaces.}
The findings also suggest that cohorts should be understood as provisional evidence spaces. 
In \toolname{}, \uline{a cohort is not a fixed category or a final answer. It is a temporary working set created by a query, context condition, motif, or emerging hypothesis.} Participants constructed cohorts, inspected aggregate patterns, revised the cohorts, and returned to individual paintings to judge whether the patterns were meaningful.
This behavior matches the iterative nature of compositional analysis. A scholar may begin with a broad question, such as the difference between Northern Song and Southern Song landscapes, or with a motif, such as a fisherman in a boat. The relevant cohort is often not known in advance. It is formed and adjusted during analysis. Therefore, visual analytics systems for art history should not treat retrieval as a one-time operation. They should support movement among query, cohort construction, overview analysis, comparison, and close inspection~\cite{pirolli2005sensemaking,heer2012interactive,shneiderman1996eyes}.


\textbf{From visual differences to art-historical interpretation.}
\uline{Although participants could readily identify compositional differences between painting cohorts using the \syscomp{Comparison View}, understanding the art-historical significance of these differences remained challenging.} As \textbf{P1} noted, visual differences alone do not explain why they emerge or whether they reflect meaningful stylistic characteristics. Similar layouts may arise from different artistic intentions, while visually different compositions may serve comparable expressive purposes. The current \syscomp{Comparison View} effectively reveals differences in entity prevalence, spatial arrangement, and relations, but it provides limited support for interpreting the historical or stylistic significance of these patterns. Future systems should therefore complement visual comparison with interpretive evidence, for example by explaining which paintings contribute most to an observed difference, linking patterns to historical context such as artists or periods, and summarizing possible factors underlying the observed compositional variation.

\subsection{Design Implications}

The expert interviews point to four design implications for composition-oriented visual analytics systems.

\textbf{Retrieval and ranking explanations.}
Systems should explain retrieval and ranking results more clearly. For scholars, this is not only about system transparency~\cite{amershi2019guidelines}. It also helps them judge whether a match can serve as evidence for further analysis. When a painting is retrieved because of partial spatial similarity, users need to know which region, entity, or relation makes the match meaningful.

\textbf{Fuzzy composition queries.}
Systems should support more abstract and fuzzy composition queries. Experts often reason about balance, density, openness, visual weight, and approximate regions. These ideas are central to compositional interpretation, but they are difficult to express through exact entity positions or fixed bounding boxes. Future query mechanisms should allow scholars to express uncertain regions, rough layout tendencies, and higher-level spatial qualities.

\textbf{Exploration history.}
Longer exploration requires history management~\cite{heer2012interactive,ragan2015characterizing}. The canvas and query provenance preserve the analytical process, but scholars need to record why a cohort was created, why a comparison was made, and how an interpretation changed over time. Future systems could treat exploration history as an analytical entity, including saved queries, cohort versions, comparison paths, notes, and links between observations and supporting paintings.

\textbf{Uncertainty and reliability cues.}
Systems should present uncertainty and reliability more explicitly~\cite{windhager2018visualization}. Cohort summaries can look persuasive, but their strength depends on cohort size, annotation quality, and the amount of supporting evidence. Showing cohort size, support counts, annotation confidence, and outliers would help scholars judge whether a pattern is stable evidence or a tentative clue.

\subsection{Limitations and Future Work}


\textbf{Corpus coverage and cohort reliability.}
Although the corpus supports composition-oriented analysis, it does not capture the full historical, regional, and stylistic diversity of \TCP{}s. As a result, some compositional patterns observed by the system may reflect the characteristics of the current corpus rather than the broader painting tradition. In addition, the reliability of analytical results depends on the size and representativeness of the retrieved cohorts. Queries returning only a small number of paintings may produce spatial summaries and cross-cohort comparisons that are disproportionately influenced by individual works. Future work will expand the corpus by incorporating paintings from a wider range of periods, subjects, and collections, while also extending \toolname{} to support user-uploaded digitized paintings and semi-automated composition annotation.


\textbf{Broader analytical support.}
While \toolname{} supports composition-oriented analysis, our study suggests opportunities to broaden its analytical scope and representation. First, experts noted that broader art-historical questions often require more flexible cohort construction than the current composition-driven queries provide, which is reflected in the mixed responses to Q3. Second, the current representation models composition through entities, spatial organization, relations, void, and context, but does not yet capture other artistic characteristics considered by art historians. For example, \toolname{} currently excludes handscrolls from spatial comparison and therefore cannot represent their sequential viewing process, inscriptions, seals, and colophons~\cite{fong1992beyond,zhang2024scrolltimes}. Experts also highlighted complementary features such as brushwork, ink usage, texture strokes, and painting style, which remain outside the current representation. Future work could extend the system by incorporating format-specific viewing logic and additional artistic characteristics.

\textbf{Evaluation scope.}
Our evaluation combines two representative case studies, a user study, and expert interviews, providing evidence for the usefulness of \toolname{} in composition-oriented analysis. The current evaluation mainly focuses on short-term tasks and representative scenarios. It involved 12 domain participants, and each session lasted about 90 minutes. Therefore, the results mainly show the usability and analytical potential of the system in a controlled setting. Future work could include longer deployments with art historians, museum researchers, and educators in real research or teaching contexts. Such studies could further examine how scholars use cohort history, comparison results, and painting-level evidence, and how system-supported observations develop into research questions, teaching materials, or art-historical arguments.

\section{Conclusion}

We present \toolname{}, a canvas-based visual analytics system for composition-oriented analysis of \TCP collections. Drawing on art-historical concepts of entities, relations, voids, and context, we propose the \compograph{}, a representation that makes painting composition explicit, queryable, and comparable. \toolname{} instantiates this representation through structure-aware retrieval, visual query, cohort construction, and coordinated views for distribution, relation, comparison, and painting-level evidence inspection. The \compograph{} serves as both a computational model for organizing compositional evidence and an analytical scaffold for scholars to move between cohort-level patterns and close reading of individual paintings. Two case studies, a user study with 12 domain participants, and expert interviews show that \toolname{} supports composition-oriented pattern discovery, cohort refinement, and evidence inspection in representative \TCP{} analysis scenarios. We hope this approach, which connects structured visual representation with expert interpretation, can support computational art history and extend to other visual-cultural collections where spatial organization is central to meaning.

\bibliographystyle{IEEEtran}
\bibliography{bibliography}

@book{jiang2012composition,
  author    = {Yue Jiang},
  title     = {Huihua Goutu yu Chuangzuo [Painting Composition and Creation]},
  publisher = {Anhui Fine Arts Publishing House},
  address   = {Hefei, China},
  year      = {2012},
  note      = {in Chinese, ISBN 978-7-5398-3231-9}
}

@article{Xue_2022, title={A study of the influence of design composition on Chinese painting}, volume={1}, url={https://drpress.org/ojs/index.php/hiaad/article/view/2075}, DOI={10.54097/hiaad.v1i2.2075}, abstractNote={
Chinese painting is one of the representatives of Chinese traditional culture and its art and cultural value is very high. By analyzing the influence of design composition on Chinese painting, this paper argues that the establishment of design composition not only enhances the art level of Chinese painting, but also has a profound influence on the thought of Chinese painting. In this regard, we need to deeply understand the characteristics of Chinese painting and strive to create Chinese painting works with unique aesthetic characteristics and personality traits on this basis.
}, number={2}, journal={Highlights in Art and Design}, author={Xue, Bing}, year={2022}, month={Oct.}, pages={58–60} }

@misc{vistcp,
      title={VisTCP: A Visualization Framework to Construct Knowledge-Graph-Based Representation for Traditional Chinese Painting}, 
      author={Zhiguang Zhou and Fengling Zheng and Miaoxin Hu and Lina You and Jin Wen and Huan Liu and Wei Zhang and Dekun Qian and Yuhua Liu and Wei Chen and Yigang Wang and Yong Wang},
      year={2026},
      eprint={2607.05841},
      archivePrefix={arXiv},
      primaryClass={cs.HC},
      url={https://arxiv.org/abs/2607.05841}, 
}

@article{Zhang_2024,
  title     = {Computational Approaches for Traditional Chinese Painting: From the “Six Principles of Painting” Perspective},
  volume    = {39},
  issn      = {1860-4749},
  url       = {http://dx.doi.org/10.1007/s11390-024-3408-x},
  doi       = {10.1007/s11390-024-3408-x},
  number    = {2},
  journal   = {Journal of Computer Science and Technology},
  publisher = {Springer Science and Business Media LLC},
  author    = {Zhang, Wei and Zhang, Jian-Wei and Wong, Kam-Kwai and Wang, Yi-Fang and Feng, Ying-Chao-Jie and Wang, Lu-Wei and Chen, Wei},
  year      = {2024},
  month     = Mar,
  pages     = {269–285}
}

@article{Munakata_1968,
  title   = {The Chinese Theory of Art. By Lin Yutang. New York: G. P. Putnam’s Sons, 1967. xii, 244 pp. 
  Illustrations, Charts, Tables, Index. \$5.95.},
  volume  = {27},
  doi     = {10.2307/2051773},
  number  = {2},
  journal = {The Journal of Asian Studies},
  author  = {Munakata, Kiyohiko},
  year    = {1968},
  pages   = {385–385}
}

@inproceedings{seguin2016visual,
  title={Visual link retrieval in a database of paintings},
  author={Seguin, Benoit and Striolo, Carlotta and diLenardo, Isabella and Kaplan, Fr{\'e}d{\'e}ric},
  booktitle={European conference on computer vision},
  pages={753--767},
  year={2016},
  organization={Springer}
}

@inproceedings{shen2019discovering,
  title={Discovering visual patterns in art collections with spatially-consistent feature learning},
  author={Shen, Xi and Efros, Alexei A and Aubry, Mathieu},
  booktitle={Proceedings of the IEEE/CVF conference on computer vision and pattern recognition},
  pages={9278--9287},
  year={2019}
}

@inproceedings{madhu2020understanding,
  title={Understanding compositional structures in art historical images using pose and gaze priors: Towards scene understanding in digital art history},
  author={Madhu, Prathmesh and Marquart, Tilman and Kosti, Ronak and Bell, Peter and Maier, Andreas and Christlein, Vincent},
  booktitle={European Conference on Computer Vision},
  pages={109--125},
  year={2020},
  organization={Springer}
}

@article{Braun01012006,
author = {Virginia Braun and Victoria Clarke},
title = {Using thematic analysis in psychology},
journal = {Qualitative Research in Psychology},
volume = {3},
number = {2},
pages = {77--101},
year = {2006},
publisher = {Routledge},
doi = {10.1191/1478088706qp063oa},
URL = {https://doi.org/10.1191/1478088706qp063oa},
eprint = {https://doi.org/10.1191/1478088706qp063oa}
}

@article{madhu2023icc++,
  title={ICC++: Explainable feature learning for art history using image compositions},
  author={Madhu, Prathmesh and Marquart, Tilman and Kosti, Ronak and Suckow, Dirk and Bell, Peter and Maier, Andreas and Christlein, Vincent},
  journal={Pattern Recognition},
  volume={136},
  pages={109153},
  year={2023},
  publisher={Elsevier}
}

@inproceedings{jenicek2019linking,
  title={Linking art through human poses},
  author={Jenicek, Tomas and Chum, Ond{\v{r}}ej},
  booktitle={2019 International Conference on Document Analysis and Recognition (ICDAR)},
  pages={1338--1345},
  year={2019},
  organization={IEEE}
}

@book{silbergeld1982chinese,
  author    = {Silbergeld, Jerome},
  title     = {Chinese Painting Style: Media, Methods, and Principles of Form},
  year      = {1982},
  publisher = {University of Washington Press},
  address   = {Seattle},
  isbn      = {0-295-95896-0}
}

@INPROCEEDINGS{dong2020,
  author={Dong, Zhenhao and Wan, Jing and Li, Chaoyue and Jiang, Han and Qian, Yingge and Pan, Wenxie},
  booktitle={2020 International Conference on Culture-oriented Science \& Technology (ICCST)}, 
  title={Feature Fusion based Cross-modal Retrieval for Traditional Chinese Painting}, 
  year={2020},
  volume={},
  number={},
  pages={383-387},
  doi={10.1109/ICCST50977.2020.00080}}

@inproceedings{blascheck2016visual,
  title={Visual analysis and coding of data-rich user behavior},
  author={Blascheck, Tanja and Beck, Fabian and Baltes, Sebastian and Ertl, Thomas and Weiskopf, Daniel},
  booktitle={2016 IEEE Conference on Visual Analytics Science and Technology (VAST)},
  pages={141--150},
  year={2016},
  organization={IEEE}
}

@article{guo2015case,
  title={A case study using visualization interaction logs and insight metrics to understand how analysts arrive at insights},
  author={Guo, Hua and Gomez, Steven R and Ziemkiewicz, Caroline and Laidlaw, David H},
  journal={IEEE transactions on visualization and computer graphics},
  volume={22},
  number={1},
  pages={51--60},
  year={2015},
  publisher={IEEE}
}

@article{reda2016modeling,
  title={Modeling and evaluating user behavior in exploratory visual analysis},
  author={Reda, Khairi and Johnson, Andrew E and Papka, Michael E and Leigh, Jason},
  journal={Information Visualization},
  volume={15},
  number={4},
  pages={325--339},
  year={2016},
  publisher={SAGE Publications Sage UK: London, England}
}

@article{liu2019qian,
  title     = {Qian Xuan’s Loyalist Revision of Iconic Imagery in Tao Yuanming Returning Home and Wang Xizhi Watching Geese},
  author    = {Liu, Shi-yee},
  journal   = {Metropolitan Museum Journal},
  volume    = {54},
  number    = {1},
  pages     = {26--46},
  year      = {2019},
  publisher = {The University of Chicago Press Chicago, IL}
}

@book{murray2007mirror,
  title     = {Mirror of morality: Chinese narrative illustration and Confucian ideology},
  author    = {Murray, Julia K},
  year      = {2007},
  publisher = {University of Hawaii Press}
}

@article{wan2024wumkg,
  title     = {WuMKG: a Chinese painting and calligraphy multimodal knowledge graph},
  author    = {Wan, Jing and Zhang, Hao and Zou, Jun and Zou, Ao and Chen, Yubin and Zeng, Qingyang and Li, Xinrong and Wang, Qiya},
  journal   = {Heritage Science},
  volume    = {12},
  number    = {1},
  pages     = {1--18},
  year      = {2024},
  publisher = {Nature Publishing Group}
}

@article{chen2025multi,
  title   = {Multi-Modal Visual Analytic Approach for Attributing and Authenticating Ancient Chinese Paintings},
  author  = {Chen, Xiaojiao and Chen, Yonghao and Tang, Tan and Ge, Xin and Wang, Ruihan and Wang, Yifan and Wang, Xiaosong},
  journal = {Journal of Computer-Aided Design \& Computer Graphics},
  volume  = {37},
  number  = {4},
  pages   = {713--724},
  year    = {2025}
}

@inproceedings{johnson2015image,
  title     = {Image retrieval using scene graphs},
  author    = {Johnson, Justin and Krishna, Ranjay and Stark, Michael and Li, Li-Jia and Shamma, David and Bernstein, Michael and Fei-Fei, Li},
  booktitle = {Proceedings of the IEEE conference on computer vision and pattern recognition},
  pages     = {3668--3678},
  year      = {2015}
}

@article{krishna2017visual,
  title={Visual genome: Connecting language and vision using crowdsourced dense image annotations},
  author={Krishna, Ranjay and Zhu, Yuke and Groth, Oliver and Johnson, Justin and Hata, Kenji and Kravitz, Joshua and Chen, Stephanie and Kalantidis, Yannis and Li, Li-Jia and Shamma, David A and others},
  journal={International journal of computer vision},
  volume={123},
  number={1},
  pages={32--73},
  year={2017},
  publisher={Springer}
}

@article{chang2021comprehensive,
  title     = {A comprehensive survey of scene graphs: Generation and application},
  author    = {Chang, Xiaojun and Ren, Pengzhen and Xu, Pengfei and Li, Zhihui and Chen, Xiaojiang and Hauptmann, Alex},
  journal   = {IEEE Transactions on Pattern Analysis and Machine Intelligence},
  volume    = {45},
  number    = {1},
  pages     = {1--26},
  year      = {2021},
  publisher = {IEEE}
}

@article{jiang2021mtffnet,
  title     = {MTFFNet: A multi-task feature fusion framework for Chinese painting classification},
  author    = {Jiang, Wei and Wang, Xiaoyu and Ren, Jinchang and Li, Sen and Sun, Meijun and Wang, Zheng and Jin, Jesse S},
  journal   = {Cognitive Computation},
  volume    = {13},
  number    = {5},
  pages     = {1287--1296},
  year      = {2021},
  publisher = {Springer}
}

@inproceedings{sun2015brushstroke,
  title        = {Brushstroke based sparse hybrid convolutional neural networks for author classification of Chinese ink-wash paintings},
  author       = {Sun, Meijun and Zhang, Dong and Ren, Jinchang and Wang, Zheng and Jin, Jesse S},
  booktitle    = {2015 IEEE International Conference on Image Processing (ICIP)},
  pages        = {626--630},
  year         = {2015},
  organization = {IEEE}
}

@article{windhager2018visualization,
  title     = {Visualization of cultural heritage collection data: State of the art and future challenges},
  author    = {Windhager, Florian and Federico, Paolo and Schreder, G{\"u}nther and Glinka, Katrin and D{\"o}rk, Marian and Miksch, Silvia and Mayr, Eva},
  journal   = {IEEE transactions on visualization and computer graphics},
  volume    = {25},
  number    = {6},
  pages     = {2311--2330},
  year      = {2018},
  publisher = {IEEE}
}

@article{glinka2017past,
  title   = {Past Visions and Reconciling Views: Visualizing Time, Texture and Themes in Cultural Collections.},
  author  = {Glinka, Katrin and Pietsch, Christopher and D{\"o}rk, Marian},
  journal = {DHQ: Digital Humanities Quarterly},
  number  = {2},
  year    = {2017}
}

@article{portales2022increasing,
  title     = {Increasing access to cultural heritage objects from multiple museums through semantically-aware maps},
  author    = {Portal{\'e}s, Cristina and Casanova-Salas, Pablo and Sevilla, Javier and Sebasti{\'a}n, Jorge and Le{\'o}n, Arabella and Samper, Jose Javier},
  journal   = {ISPRS International Journal of Geo-Information},
  volume    = {11},
  number    = {4},
  pages     = {266},
  year      = {2022},
  publisher = {MDPI}
}

@article{li2025musekg,
  title   = {MUSEKG: A Knowledge Graph Over Museum Collections},
  author  = {Li, Jinhao and Qi, Jianzhong and Han, Soyeon Caren and Holden, Eun-Jung},
  journal = {arXiv preprint arXiv:2511.16014},
  year    = {2025}
}

@article{zhang2024scrolltimes,
  title     = {Scrolltimes: Tracing the provenance of paintings as a window into history},
  author    = {Zhang, Wei and Kam-Kwai, Wong and Chen, Yitian and Jia, Ailing and Wang, Luwei and Zhang, Jian-Wei and Cheng, Lechao and Qu, Huamin and Chen, Wei},
  journal   = {IEEE Transactions on Visualization and Computer Graphics},
  volume    = {30},
  number    = {6},
  pages     = {2981--2994},
  year      = {2024},
  publisher = {IEEE}
}

@inproceedings{oda2025artevoviewer,
  title        = {ArtEvoViewer: A System for Visualizing Interpersonal Influence Among Painters},
  author       = {Oda, Ryoko and Nakamura, Eita and Pahr, Daniel and Ehlers, Henry and Gr{\"o}ller, Eduard and Raidou, Renata G and Itoh, Takayuki},
  booktitle    = {2025 29th International Conference Information Visualisation (IV)},
  pages        = {171--176},
  year         = {2025},
  organization = {IEEE}
}

@article{fan2020visual,
  title     = {Visual order of Chinese ink paintings},
  author    = {Fan, Zhen-Bao and Zhang, Kang},
  journal   = {Visual computing for industry, biomedicine, and art},
  volume    = {3},
  number    = {1},
  pages     = {23},
  year      = {2020},
  publisher = {Springer}
}

@article{feng2022ipoet,
  title     = {iPoet: interactive painting poetry creation with visual multimodal analysis},
  author    = {Feng, Yingchaojie and Chen, Jiazhou and Huang, Keyu and Wong, Jason K and Ye, Hui and Zhang, Wei and Zhu, Rongchen and Luo, Xiaonan and Chen, Wei},
  journal   = {Journal of Visualization},
  volume    = {25},
  number    = {3},
  pages     = {671--685},
  year      = {2022},
  publisher = {Springer}
}

@book{bush2012early,
  title     = {Early Chinese texts on painting},
  author    = {Bush, Susan and Shih, Hsio-yen},
  volume    = {1},
  year      = {2012},
  publisher = {Hong Kong University Press}
}

@book{arnheim1954art,
  title     = {Art and visual perception: A psychology of the creative eye},
  author    = {Arnheim, Rudolf},
  year      = {1954},
  publisher = {Univ of California Press}
}

@inproceedings{yoon2021image,
  title     = {Image-to-image retrieval by learning similarity between scene graphs},
  author    = {Yoon, Sangwoong and Kang, Woo Young and Jeon, Sungwook and Lee, SeongEun and Han, Changjin and Park, Jonghun and Kim, Eun-Sol},
  booktitle = {Proceedings of the AAAI Conference on Artificial Intelligence},
  volume    = {35},
  number    = {12},
  pages     = {10718--10726},
  year      = {2021}
}

@article{furuta2019efficient,
  title     = {Efficient and interactive spatial-semantic image retrieval},
  author    = {Furuta, Ryosuke and Inoue, Naoto and Yamasaki, Toshihiko},
  journal   = {Multimedia Tools and Applications},
  volume    = {78},
  number    = {13},
  pages     = {18713--18733},
  year      = {2019},
  publisher = {Springer}
}

@article{braun2006using,
  title     = {Using thematic analysis in psychology},
  author    = {Braun, Virginia and Clarke, Victoria},
  journal   = {Qualitative research in psychology},
  volume    = {3},
  number    = {2},
  pages     = {77--101},
  year      = {2006},
  publisher = {Taylor \& Francis}
}

@article{bostock2011d3,
  title     = {D$^3$ data-driven documents},
  author    = {Bostock, Michael and Ogievetsky, Vadim and Heer, Jeffrey},
  journal   = {IEEE transactions on visualization and computer graphics},
  volume    = {17},
  number    = {12},
  pages     = {2301--2309},
  year      = {2011},
  publisher = {IEEE}
}

@article{zeileis2009escaping,
  title     = {Escaping RGBland: Selecting colors for statistical graphics},
  author    = {Zeileis, Achim and Hornik, Kurt and Murrell, Paul},
  journal   = {Computational Statistics \& Data Analysis},
  volume    = {53},
  number    = {9},
  pages     = {3259--3270},
  year      = {2009},
  publisher = {Elsevier}
}

@misc{vueflow2025,
  author = {Burak Cakmakoglu},
  title  = {Vue Flow: A highly customizable Vue 3 Flowchart component},
  year   = {2025},
  url    = {https://github.com/bcakmakoglu/vue-flow}
}

@misc{caod2021,
  author       = {{Chinese Treasures Museum}},
  title        = {{Chinese Art Open Dataset (CAOD)}},
  year         = {2021},
  url          = {https://github.com/ltfc/CAOD/tree/main},
  note         = {Accessed: 2026-07-09}
}

@book{cahill1994painter,
  title     = {The painter's practice: How artists lived and worked in traditional China},
  author    = {Cahill, James},
  number    = {29},
  year      = {1994},
  publisher = {Columbia University Press}
}

@book{fong1992beyond,
  title     = {Beyond representation: Chinese painting and calligraphy, 8th-14th century},
  author    = {Fong, Wen},
  year      = {1992},
  publisher = {Metropolitan Museum of Art}
}

@book{manovich2020cultural,
  title     = {Cultural analytics},
  author    = {Manovich, Lev},
  year      = {2020},
  publisher = {Mit Press}
}

@article{arnold2019distant,
  title     = {Distant viewing: analyzing large visual corpora},
  author    = {Arnold, Taylor and Tilton, Lauren},
  journal   = {Digital Scholarship in the Humanities},
  volume    = {34},
  number    = {Supplement\_1},
  pages     = {i3--i16},
  year      = {2019},
  publisher = {Oxford University Press}
}

@article{felix2017taking,
  title={Taking word clouds apart: An empirical investigation of the design space for keyword summaries},
  author={Felix, Cristian and Franconeri, Steven and Bertini, Enrico},
  journal={IEEE transactions on visualization and computer graphics},
  volume={24},
  number={1},
  pages={657--666},
  year={2017},
  publisher={IEEE}
}

@inproceedings{pirolli2005sensemaking,
  title={The sensemaking process and leverage points for analyst technology as identified through cognitive task analysis},
  author={Pirolli, Peter and Card, Stuart},
  booktitle={Proceedings of international conference on intelligence analysis},
  volume={5},
  number={1},
  pages={2--4},
  year={2005},
  organization={McLean, VA, USA}
}

@article{heer2012interactive,
  title={Interactive dynamics for visual analysis: A taxonomy of tools that support the fluent and flexible use of visualizations},
  author={Heer, Jeffrey and Shneiderman, Ben},
  journal={Queue},
  volume={10},
  number={2},
  pages={30--55},
  year={2012},
  publisher={ACM New York, NY, USA}
}

@article{ragan2015characterizing,
  title={Characterizing provenance in visualization and data analysis: an organizational framework of provenance types and purposes},
  author={Ragan, Eric D and Endert, Alex and Sanyal, Jibonananda and Chen, Jian},
  journal={IEEE transactions on visualization and computer graphics},
  volume={22},
  number={1},
  pages={31--40},
  year={2015},
  publisher={IEEE}
}

@inproceedings{amershi2019guidelines,
  title={Guidelines for human-AI interaction},
  author={Amershi, Saleema and Weld, Dan and Vorvoreanu, Mihaela and Fourney, Adam and Nushi, Besmira and Collisson, Penny and Suh, Jina and Iqbal, Shamsi and Bennett, Paul N and Inkpen, Kori and others},
  booktitle={Proceedings of the 2019 chi conference on human factors in computing systems},
  pages={1--13},
  year={2019}
}

@inproceedings{Ye2025HyperMOOCAM,
author = {Ye, Li and Wang, Lei and Cai, Lihong and Yu, Ruiqi and Wang, Yong and Wang, Yigang and Chen, Wei and Zhou, Zhiguang},
title = {HyperMOOC: Augmenting MOOC Videos with Concept-based Embedded Visualizations},
year = {2026},
isbn = {9798400722783},
publisher = {Association for Computing Machinery},
address = {New York, NY, USA},
url = {https://doi.org/10.1145/3772318.3791377},
doi = {10.1145/3772318.3791377},
booktitle = {Proceedings of the 2026 CHI Conference on Human Factors in Computing Systems},
articleno = {1609},
numpages = {18},
location = {
},
series = {CHI '26}
}

@article{xiao2026square,
  title={SQUARE FULL COMPOSITION IN CONTEMPORARY CHINESE FIGURE PAINTING: SPATIAL DENSITY AND VISUAL NARRATIVE IN THE 12TH--14TH NATIONAL ART EXHIBITIONS},
  author={Xiao, Tianwei and binti Kindoyop, Salbiah and Sahibil, Zaimie Bin},
  journal={Jurnal Gendang Alam (GA)},
  volume={16},
  number={1},
  year={2026}
}

@book{hearn2008read,
  title={How to read Chinese paintings},
  author={Hearn, Maxwell K},
  year={2008},
  publisher={Metropolitan Museum of Art}
}

@book{barnhart1997three,
  title={Three thousand years of Chinese painting},
  author={Barnhart, Richard M and Yang, Xin and Chongzheng, Nie and Cahill, James and Wu, Hung and Shaojun, Lang},
  year={1997},
  publisher={Yale University Press}
}

@book{fong1973sung,
  title={Sung and Yuan paintings},
  author={Fong, Wen and Fu, Marilyn},
  year={1973},
  publisher={Metropolitan museum of art}
}

@book{cahill1976hills,
  title={Hills beyond a river: Chinese painting of the Y{\"u}an Dynasty, 1279-1368},
  author={Cahill, James},
  year={1976},
  publisher={Weatherhill}
}

@book{murck1991words,
  title={Words and images: Chinese poetry, calligraphy, and painting},
  author={Murck, Alfreda and Fong, Wen},
  year={1991},
  publisher={Metropolitan Museum of Art}
}

@article{delbanco2000chinese,
  title={Chinese handscrolls},
  author={Delbanco, Dawn},
  journal={Heilbrunn Timeline of Art History. New York: The Metropolitan Museum of Art},
  year={2000}
}

@article{krzywinski2009circos,
  title={Circos: an information aesthetic for comparative genomics},
  author={Krzywinski, Martin and Schein, Jacqueline and Birol, Inanc and Connors, Joseph and Gascoyne, Randy and Horsman, Doug and Jones, Steven J and Marra, Marco A},
  journal={Genome research},
  volume={19},
  number={9},
  pages={1639--1645},
  year={2009},
  publisher={Cold Spring Harbor Lab}
}

@article{wongsuphasawat2015voyager,
  title={Voyager: Exploratory analysis via faceted browsing of visualization recommendations},
  author={Wongsuphasawat, Kanit and Moritz, Dominik and Anand, Anushka and Mackinlay, Jock and Howe, Bill and Heer, Jeffrey},
  journal={IEEE transactions on visualization and computer graphics},
  volume={22},
  number={1},
  pages={649--658},
  year={2015},
  publisher={IEEE}
}

@inproceedings{ahlberg1992dynamic,
  title={Dynamic queries for information exploration: An implementation and evaluation},
  author={Ahlberg, Christopher and Williamson, Christopher and Shneiderman, Ben},
  booktitle={Proceedings of the SIGCHI conference on Human factors in computing systems},
  pages={619--626},
  year={1992}
}

@article{shneiderman2002dynamic,
  title={Dynamic queries for visual information seeking},
  author={Shneiderman, Ben},
  journal={IEEE software},
  volume={11},
  number={6},
  pages={70--77},
  year={2002},
  publisher={IEEE}
}

@article{stolte2002,
author = {Stolte, Chris and Tang, Diane and Hanrahan, Pat},
title = {Polaris: A System for Query, Analysis, and Visualization of Multidimensional Relational Databases},
year = {2002},
issue_date = {January 2002},
publisher = {IEEE Educational Activities Department},
address = {USA},
volume = {8},
number = {1},
issn = {1077-2626},
url = {https://doi.org/10.1109/2945.981851},
doi = {10.1109/2945.981851},
journal = {IEEE Transactions on Visualization and Computer Graphics},
month = jan,
pages = {52–65},
numpages = {14}
}

@article{fan2019,
    author = {Fan, ZhenBao and Zhang, Kang and Zheng, XianJun Sam},
    title = {Evaluation and Analysis of White Space in Wu Guanzhong’s Chinese Paintings},
    journal = {Leonardo},
    volume = {52},
    number = {2},
    pages = {111-116},
    year = {2019},
    month = {04},
    issn = {0024-094X},
    doi = {10.1162/leon_a_01409},
    url = {https://doi.org/10.1162/leon_a_01409},
    eprint = {https://direct.mit.edu/leon/article-pdf/52/2/111/1578493/leon_a_01409.pdf},
}

@book{gonzalez2009digital,
  title={Digital image processing},
  author={Gonzalez, Rafael C},
  year={2009},
  publisher={Pearson education india}
}

@article{sobel19683x3,
  title={A 3x3 isotropic gradient operator for image processing},
  author={Sobel, Irwin and Feldman, Gary and others},
  journal={a talk at the Stanford Artificial Project in},
  volume={1968},
  pages={271--272},
  year={1968}
}

@article{kong2015,
author = {Kong, Lingfu and Duan, Liangliang and Yang, Wenji and Dou, Yan},
title = {Salient region detection: an integration approach based on image pyramid and region property},
year = {2015},
issue_date = {February 2015},
publisher = {John Wiley \& Sons, Inc.},
address = {USA},
volume = {9},
number = {1},
url = {https://doi.org/10.1049/iet-cvi.2013.0285},
doi = {10.1049/iet-cvi.2013.0285},
journal = {IET Computer Vision},
month = feb,
pages = {85–97},
numpages = {13}
}

@article{he2008run,
  title={A run-based two-scan labeling algorithm},
  author={He, Lifeng and Chao, Yuyan and Suzuki, Kenji},
  journal={IEEE transactions on image processing},
  volume={17},
  number={5},
  pages={749--756},
  year={2008},
  publisher={IEEE}
}

@article{snyder2019literature,
    title = {Literature review as a research methodology: An overview and guidelines},
    journal = {Journal of Business Research},
    volume = {104},
    pages = {333-339},
    year = {2019},
    issn = {0148-2963},
    doi = {https://doi.org/10.1016/j.jbusres.2019.07.039},
    url = {https://www.sciencedirect.com/science/article/pii/S0148296319304564},
    author = {Hannah Snyder}
}

@article{okoli2015guide,
    author = {Okoli, Chitu},
    year = {2015},
    month = {11},
    pages = {},
    title = {A Guide to Conducting a Standalone Systematic Literature Review},
    volume = {37},
    journal = {Communications of the Association for Information Systems},
    doi = {10.17705/1CAIS.03743}
}

@article{kitchenham2007guidelines,
      title={Guidelines for performing systematic literature reviews in software engineering},
      author={Kitchenham, Barbara and Charters, Stuart and others},
      year={2007},
      publisher={Keele, UK}
}

@inproceedings{wohlin2014guidelines,
    author = {Wohlin, Claes},
    title = {Guidelines for snowballing in systematic literature studies and a replication in software engineering},
    year = {2014},
    isbn = {9781450324762},
    publisher = {Association for Computing Machinery},
    address = {New York, NY, USA},
    url = {https://doi.org/10.1145/2601248.2601268},
    doi = {10.1145/2601248.2601268},
    booktitle = {Proceedings of the 18th International Conference on Evaluation and Assessment in Software Engineering},
    articleno = {38},
    numpages = {10},
    location = {London, England, United Kingdom},
    series = {EASE '14}
}

@article{haddaway2015role,
  title={The role of Google Scholar in evidence reviews and its applicability to grey literature searching},
  author={Haddaway, Neal Robert and Collins, Alexandra Mary and Coughlin, Deborah and Kirk, Stuart},
  journal={PloS one},
  volume={10},
  number={9},
  pages={e0138237},
  year={2015},
  publisher={Public Library of Science San Francisco, CA USA}
}

@article{templier2015framework,
  title={A framework for guiding and evaluating literature reviews},
  author={Templier, Mathieu and Par{\'e}, Guy},
  journal={Communications of the Association for Information Systems},
  volume={37},
  number={1},
  pages={6},
  year={2015}
}

@inproceedings{baldonado2000guidelines,
    author = {Wang Baldonado, Michelle Q. and Woodruff, Allison and Kuchinsky, Allan},
    title = {Guidelines for using multiple views in information visualization},
    year = {2000},
    isbn = {1581132522},
    publisher = {Association for Computing Machinery},
    address = {New York, NY, USA},
    url = {https://doi.org/10.1145/345513.345271},
    doi = {10.1145/345513.345271},
    booktitle = {Proceedings of the Working Conference on Advanced Visual Interfaces},
    pages = {110–119},
    numpages = {10},
    location = {Palermo, Italy},
    series = {AVI '00}
}

@article{gleicher2011visual,
  title={Visual comparison for information visualization},
  author={Gleicher, Michael and Albers, Danielle and Walker, Rick and Jusufi, Ilir and Hansen, Charles D and Roberts, Jonathan C},
  journal={Information Visualization},
  volume={10},
  number={4},
  pages={289--309},
  year={2011},
  publisher={SAGE Publications Sage UK: London, England}
}

@inproceedings{shneiderman1996eyes,
  author={Shneiderman, B.},
  booktitle={Proceedings 1996 IEEE Symposium on Visual Languages}, 
  title={The eyes have it: a task by data type taxonomy for information visualizations}, 
  year={1996},
  volume={},
  number={},
  pages={336-343},
  doi={10.1109/VL.1996.545307}
}

@article{cleveland1984graphical,
 ISSN = {01621459, 1537274X},
 URL = {http://www.jstor.org/stable/2288400},
 author = {William S. Cleveland and Robert McGill},
 journal = {Journal of the American Statistical Association},
 number = {387},
 pages = {531--554},
 publisher = {[American Statistical Association, Taylor & Francis, Ltd.]},
 title = {Graphical Perception: Theory, Experimentation, and Application to the Development of Graphical Methods},
 urldate = {2026-07-25},
 volume = {79},
 year = {1984}
}

@article{misue1995layout,
  title={Layout adjustment and the mental map},
  author={Misue, Kazuo and Eades, Peter and Lai, Wei and Sugiyama, Kozo},
  journal={Journal of Visual Languages \& Computing},
  volume={6},
  number={2},
  pages={183--210},
  year={1995},
  publisher={Elsevier}
}

@inproceedings{lan2012structured,
    author = {Lan, Tian and Yang, Weilong and Wang, Yang and Mori, Greg},
    title = {Image retrieval with structured object queries using latent ranking SVM},
    year = {2012},
    isbn = {9783642337826},
    publisher = {Springer-Verlag},
    address = {Berlin, Heidelberg},
    url = {https://doi.org/10.1007/978-3-642-33783-3_10},
    doi = {10.1007/978-3-642-33783-3_10},
    booktitle = {Proceedings of the 12th European Conference on Computer Vision - Volume Part VI},
    pages = {129–142},
    numpages = {14},
    location = {Florence, Italy},
    series = {ECCV'12}
}

@inproceedings{pienta2016visage,
    author = {Pienta, Robert and Tamersoy, Acar and Endert, Alex and Navathe, Shamkant and Tong, Hanghang and Chau, Duen Horng},
    title = {VISAGE: Interactive Visual Graph Querying},
    year = {2016},
    isbn = {9781450341318},
    publisher = {Association for Computing Machinery},
    address = {New York, NY, USA},
    url = {https://doi.org/10.1145/2909132.2909246},
    doi = {10.1145/2909132.2909246},
    booktitle = {Proceedings of the International Working Conference on Advanced Visual Interfaces},
    pages = {272–279},
    numpages = {8},
    location = {Bari, Italy},
    series = {AVI '16}
}

@article{glueck2017phenostacks,
  author={Glueck, Michael and Gvozdik, Alina and Chevalier, Fanny and Khan, Azam and Brudno, Michael and Wigdor, Daniel},
  journal={IEEE Transactions on Visualization and Computer Graphics}, 
  title={PhenoStacks: Cross-Sectional Cohort Phenotype Comparison Visualizations}, 
  year={2017},
  volume={23},
  number={1},
  pages={191-200},
  doi={10.1109/TVCG.2016.2598469}
}

@inproceedings{agrawal1993mining,
    author = {Agrawal, Rakesh and Imieli\'{n}ski, Tomasz and Swami, Arun},
    title = {Mining association rules between sets of items in large databases},
    year = {1993},
    isbn = {0897915925},
    publisher = {Association for Computing Machinery},
    address = {New York, NY, USA},
    url = {https://doi.org/10.1145/170035.170072},
    doi = {10.1145/170035.170072},
    booktitle = {Proceedings of the 1993 ACM SIGMOD International Conference on Management of Data},
    pages = {207–216},
    numpages = {10},
    location = {Washington, D.C., USA},
    series = {SIGMOD '93}
}

@article{deng2023visimages,
    author = {Deng, Dazhen and Wu, Yihong and Shu, Xinhuan and Wu, Jiang and Fu, Siwei and Cui, Weiwei and Wu, Yingcai},
    title = {VisImages: A Fine-Grained Expert-Annotated Visualization Dataset},
    year = {2023},
    issue_date = {July 2023},
    publisher = {IEEE Educational Activities Department},
    address = {USA},
    volume = {29},
    number = {7},
    issn = {1077-2626},
    url = {https://doi.org/10.1109/TVCG.2022.3155440},
    doi = {10.1109/TVCG.2022.3155440},
    journal = {IEEE Transactions on Visualization and Computer Graphics},
    month = jul,
    pages = {3298–3311},
    numpages = {14}
}

@article{somarakis2021visual,
  author={Somarakis, Antonios and Ijsselsteijn, Marieke E. and Luk, Sietse J. and Kenkhuis, Boyd and de Miranda, Noel F.C.C. and Lelieveldt, Boudewijn P.F. and Höllt, Thomas},
  journal={IEEE Transactions on Visualization and Computer Graphics}, 
  title={Visual cohort comparison for spatial single-cell omics-data}, 
  year={2021},
  volume={27},
  number={2},
  pages={733-743},
  doi={10.1109/TVCG.2020.3030336}
}

\vfill

\end{document}